\documentclass[letterpaper]{article} % DO NOT CHANGE THIS
\usepackage[draft]{aaai2026}  % DO NOT CHANGE THIS
\usepackage{times}  % DO NOT CHANGE THIS
\usepackage{helvet}  % DO NOT CHANGE THIS
\usepackage{courier}  % DO NOT CHANGE THIS
\usepackage[hyphens]{url}  % DO NOT CHANGE THIS
\usepackage{graphicx} % DO NOT CHANGE THIS
\usepackage{natbib}  % DO NOT CHANGE THIS AND DO NOT ADD ANY OPTIONS TO IT
\usepackage{caption} % DO NOT CHANGE THIS AND DO NOT ADD ANY OPTIONS TO IT
\usepackage{todonotes}

\usepackage{algorithm}
\usepackage{algorithmic}

\usepackage{newfloat}
\usepackage{listings}
\DeclareCaptionStyle{ruled}{labelfont=normalfont,labelsep=colon,strut=off} % DO NOT CHANGE THIS
\floatstyle{ruled}
\newfloat{listing}{tb}{lst}{}
\floatname{listing}{Listing}
\usepackage{amsmath,amssymb,amsfonts}
\usepackage{amsthm}
\usepackage{mathtools}
\usepackage{stmaryrd}
\usepackage{bbold}
\usepackage{subcaption}
\usepackage{stmaryrd}
\usepackage{pifont}
\usepackage{cleveref}
\usepackage{booktabs}
\usepackage{pdfpages}

\usepackage[acronym]{glossaries}
\newacronym{ctde}{CTDE}{Centralized Training with Decentralized Execution}
\newacronym{cte}{CTE}{Centralized Training and Execution}
\newacronym{cpe}{CPE}{Centralized Permutation Equivariant}
\newacronym{marl}{MARL}{Multi-Agent Reinforcement Learning}
\newacronym{drl}{DRL}{Deep Reinforcement Learning}
\newacronym{mpe}{MPE}{Multi-Agent Particle Environment}
\newacronym{rware}{RWARE}{multi-robot warehouse task}
\newacronym{smac}{SMAC}{ StarCraft Multi-Agent Challenge}
\newacronym{dec-pomdp}{Dec-POMDP}{Decentralized Partially Observable Markov Decision Process}
\newacronym{rnn}{RNN}{Recurrent Neural Networks}
\newacronym{rl}{RL}{Reinforcement Learning}
\newacronym{pe}{PE}{Permutation Equivariant}
\newacronym{pi}{PI}{Permutation Invariant}
\newacronym{glpe}{GLPE}{Global-Local PE}
\newacronym{mlp}{MLP}{Multilayer Perceptron}

\newcommand{\sublayer}{g}
\newcommand{\sublayerfunc}{v}

\newtheorem{definition}{Definition}

\newtheorem{proposition}{Proposition}[section]

\crefname{lemma}{lemma}{lemmas}
\Crefname{lemma}{Lemma}{Lemmas}
\crefname{theorem}{theorem}{theorems}
\Crefname{theorem}{Theorem}{Theorems}
\crefname{proposition}{proposition}{propositions}
\Crefname{proposition}{Proposition}{Propositions}

\title{Centralized Permutation Equivariant Policy for Cooperative Multi-Agent Reinforcement Learning}
\author{
    Zhuofan Xu\textsuperscript{\rm 1}, Benedikt Bollig\textsuperscript{\rm 1}, Matthias Függer\textsuperscript{\rm 1}, Thomas Nowak\textsuperscript{\rm 1,}\textsuperscript{\rm 2} and Vincent Le Dréau\textsuperscript{\rm 1}
}
\affiliations{
    \textsuperscript{\rm 1}Université Paris-Saclay,  CNRS, ENS Paris-Saclay, LMF, Gif-sur-Yvette, France\\
    \textsuperscript{\rm 2}Institut Universitaire de France, Gif-sur-Yvette, France
}

\begin{document}

\maketitle
\begin{abstract}
The Centralized Training with Decentralized Execution (CTDE) paradigm has become increasingly popular in the multi-agent reinforcement learning community and is widely adopted in recent works.
However, decentralized policies work with partial observations and may achieve suboptimal reward when compared to centralized policies, while centralized policies may not be scalable to higher number of agents.

To address these limitations, we introduce Centralized Permutation Equivariant (CPE) learning, a centralized training and execution framework that employs a fully centralized policy to enhance the performance of standard CTDE algorithms.
Our policy network is built upon a novel permutation equivariant architecture, Global--Local Permutation Equivariant networks, which are lightweight, scalable, and easy to implement.
Empirical results demonstrate that CPE can be seamlessly integrated with both value decomposition and actor--critic methods, significantly improving performance of classical CTDE methods across cooperative benchmarks such as MPE, SMAC, and RWARE, and meet performance of state-of-the-art RWARE implementations.
\end{abstract}


\section{Introduction}
\label{sec:intro}

Multi-Agent Reinforcement Learning (MARL) has received increasing attention from the Deep Reinforcement Learning (DRL) community. Its applications span various domains where multi-agent systems are required, such as traffic control~\cite{art:traffic} and games~\cite{art:game,art:hpn}. However, MARL inherits the curse of dimensionality from DRL, and this issue becomes even more severe in the multi-agent setting. When joint observations and joint actions are constructed from individual agents' observations and actions, the size of the overall observation and action spaces grows exponentially with the number of agents. This imposes a significant burden on centralized approaches, which rely on mapping joint observations to joint actions.

% MARL and CTDE
Recently, a class of MARL algorithms has emerged under the \gls{ctde} paradigm, such as QMIX~\cite{art:qmix}, QPLEX~\cite{art:qplex}, and MAPPO~\cite{art:mappo}, showing strong performance across various MARL benchmarks. 
The core idea of \gls{ctde} is to use decentralized policies for execution while incorporating centralized components during training to enhance coordination and learning efficiency. 
\gls{ctde} methods commonly share policy parameters among agents, further improving sample efficiency and reducing the number of trainable parameters.

\paragraph{Limits of CTDE.}
Despite the strong empirical performance of \gls{ctde} methods, it remains unclear whether decentralization consistently improves policy quality. Two common arguments in favor of decentralized policies are: (1) grounding decisions in local observations avoids the exponential growth of observation and action spaces, and (2) it reflects real-world scenarios where inter-agent communication is restricted. However, these assumptions are not always substantiated by current MARL benchmarks. In many widely used testbeds, the number of agents is relatively small (often fewer than five), making scalability a less critical issue. Moreover, in tasks such as \gls{mpe}, agents often observe full state information about others (e.g., positions and velocities), contradicting assumptions about either limited communication or partial observability.

\begin{figure}[bth]
    \centering
    \includegraphics[width=0.5\linewidth]{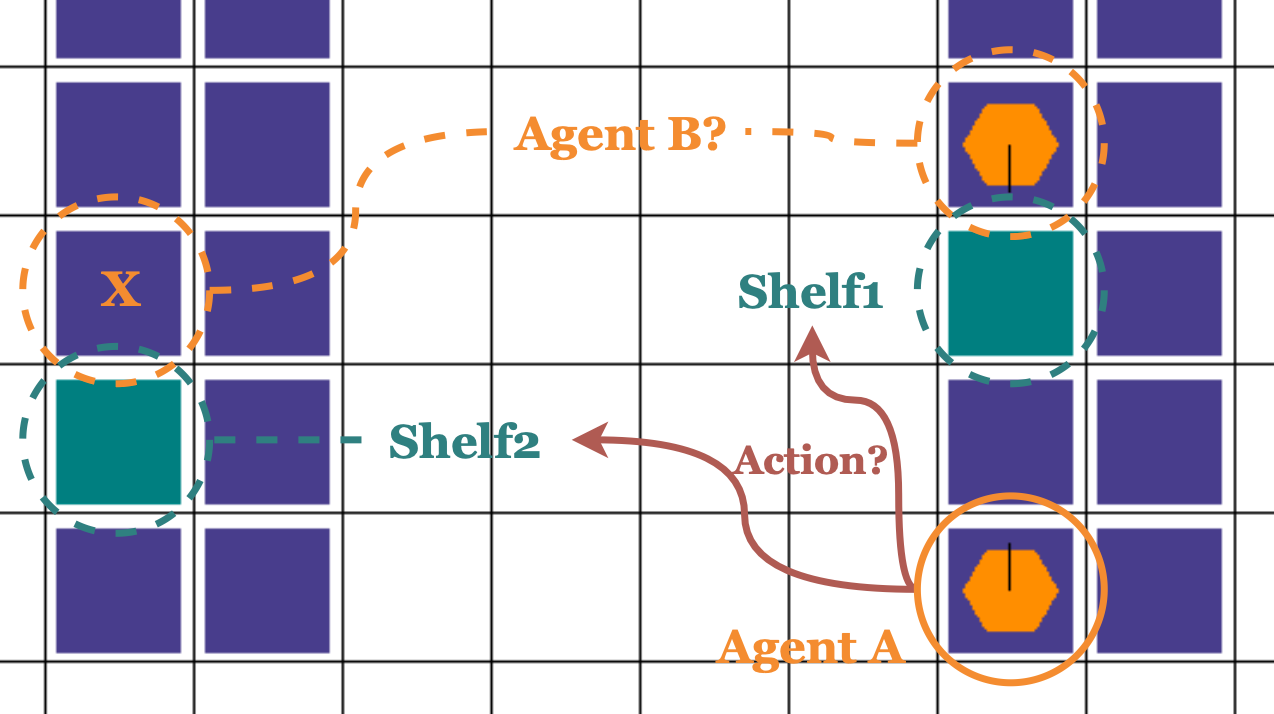}
    \caption{A state in \gls{rware} where a policy based only on local observations potentially leads to suboptimal decisions.}
    \label{fig:rwareexamp}
\end{figure}

A more critical concern with decentralized policies is that, in certain situations, local observations may lack sufficient information for optimal decision-making. For example, consider a state in the \gls{rware} shown in~\Cref{fig:rwareexamp}, where robots (agents) (orange hexagons) are required to collect designated shelves (green squares). Agent A is about to choose its next action. If Agent B’s position is as shown in the figure, the optimal decision for Agent A would be to collect Shelf2; but if Agent B is instead at position X, Agent A should collect Shelf1. If Agent A’s observation includes only its own position and the positions of the shelves, without awareness of other agents’ locations, it cannot distinguish between these two scenarios and is likely to make suboptimal choices.

\paragraph{Contribution: scalable centralized network.}
Motivated by these limitations, we investigate whether centralized policies can outperform decentralized ones in cooperative settings. To this end, we propose \textbf{Centralized Permutation Equivariant learning (CPE)}, a framework that transforms any \gls{ctde} algorithm into a \gls{cte} paradigm. CPE retains the centralized training components of the original \gls{ctde} methods, such as value decomposition or centralized critics, thereby preserving their ability to recondition local rewards and leverage additional state information during training, while replacing decentralized execution with a fully centralized one.

For the network that outputs Q-values or policies, we design a structure, termed the \textbf{Global-Local Permutation Equivariant (GLPE) network}, that abstracts global information from local observations and re-integrates it with local features. To ensure scalability and preserve input structural information, GLPE is inherently permutation equivariant. We show that such networks are agnostic to the number of agents, meaning their structure does not significantly change as the agent count varies. As a result, the model remains lightweight and scalable, with its complexity growing gracefully, or even remaining constant when the individual observation space is fixed.

% TODO: Adjust this part about experimental results
We evaluated our method on \gls{mpe}, for which we propose more challenging settings where agents' observations are strictly local, \gls{smac} and \gls{rware}. Our CPE method leads to substantial performance improvements on variant scenarios of these tasks, and achieves state-of-the-art results on \gls{rware}. The proposed architecture also demonstrates strong scalability: the policy network\footnote{Precisely, the network predicting the action or the distribution of actions is the policy network. Since in the value based method, the value function determines the action indirectly, for simplicity, we use the term policy network also for value functions.} remains compact as the number of agents increases. We also showcase the impact of joint observations on policy quality. These findings suggest that centralized policies still hold significant advantages and merit further investigation by the MARL community.

\section{MARL, CTDE, and PE}

We begin with notation and a brief review of key concepts.

\paragraph{MARL and Dec-POMDP.}
Multi-Agent Reinforcement Learning (MARL) is a subfield of reinforcement learning that considers environments with $N > 1$ agents. In fully cooperative MARL tasks, agents work together to maximize the cumulative reward. Such tasks are commonly modeled as a Decentralized Partially Observable Markov Decision Process (Dec-POMDP)~\cite{art:decpomdp2}, defined by the tuple $\langle S, A, P, r, Z, O, N, \gamma \rangle$.

Here, $s \in S$ denotes the global state of the environment, which is not directly accessible to individual agents. Due to partial observability, at each timestep, every agent $i \in D := \{1, \dots, N\}$ receives an individual observation $z_i \in Z_i$ according to the observation function $O_i(s): S \to Z_i$. The joint observation consists of all agents' observations, written as $O(s) = (O_i(s))_{i \in D}$, which is contained in $Z = \prod_{i \in D} Z_i$.
For the joint action space we write $A = \prod_{i \in D} A_i$, where $a_i \in A_i$ is the action of agent $i$. The environment transitions to the next state according to the transition function $P(s' \mid s, a): S \times A \times S \to [0,1]$. A shared reward is assigned to all agents by the reward function $r(s, a): S \times A \to \mathbb{R}$, in the cooperative setting. The cumulative reward is discounted by a factor $\gamma \in [0, 1)$.
Individual agent policies are denoted by $\pi_{i, \theta_i}(a_i \mid z_i): Z_i \times A_i \to [0, 1]$, where $\theta_i$ represents the parameters of agent $i$'s policy. The joint policy is denoted by $\pi_{\theta}$, parameterized by the set $\theta = \{\theta_i\}_{i=1}^N$.
Many MARL algorithms employ Recurrent Neural Networks (RNNs), making the policy dependent not only on the current observation $z_i$, but also on the hidden state $h_i \in H_i$ of the agent. In such cases, the policy becomes $\pi_{i, \theta_i}(a_i \mid z_i, h_i): Z_i \times H_i \times A_i \to [0, 1]$.

\paragraph{CTDE.}
Most MARL algorithms build on single-agent RL methods like Deep Q-Learning~\cite{art:dqn} and Actor-Critic~\cite{art:a3c,art:ppo}, but face scalability and instability challenges due to the exponential growth of joint spaces and non-stationarity from other agents. The \gls{ctde} paradigm addresses these issues by maintaining decentralized policies while using centralized training components (e.g., critics or mixing networks) that leverage global state information to enhance coordination.

% \paragraph{Value Mixing.} 
Value-based \gls{ctde} methods assign each agent a local Q-function $Q_{\theta_i}(z_i, a_i)$ for execution. During training, these are combined into a global Q-value $Q_{\text{tot}}(z, a)$ optimized with the global reward.
\textbf{VDN}~\cite{art:vdn} approximates $Q_{\text{tot}}$ as the sum of local Q-values,
% $Q_{\text{tot}}(z, a) \approx \sum_i Q_i(z_i, a_i).$
\textbf{QMIX}~\cite{art:qmix} extends VDN with a monotonic mixing network conditioned on the global state.
% enforcing $\partial Q_{\text{tot}} / \partial Q_i \geq 0$.
\textbf{QPLEX}~\cite{art:qplex} introduces a dueling-style decomposition, splitting each $Q_i$ into a value $V_i$ and advantage $Ad_i$, which are separately mixed using state and joint actions.
% enabling better approximation of complex interactions.

% \paragraph{Centralized Critic.} 
Actor-Critic-based \gls{ctde} methods assign local actors to agents and use a centralized critic during training to estimate value functions.
% with access to global state and joint actions.
\textbf{MAA2C}~\cite{art:maa2c} (Central-V) extends A2C~\cite{art:a3c} with a centralized critic over the global state instead of local histories.
\textbf{MAPPO}~\cite{art:mappo} adapts PPO~\cite{art:ppo} for multi-agent settings by pairing local policies with a shared centralized critic.
% performing well in \gls{mpe}~\cite{art:pymarlzooplus}.

\subsection{Permutation Equivariance}
\label{sec:preliminarype}

In many MARL tasks, agent identities are interchangeable, making input ordering arbitrary. In such cases, functions are expected to produce outputs that either permute in the same way as the inputs (\emph{\gls{pe}}), or remain unchanged regardless of input order (\emph{\gls{pi}})\footnote{We use \gls{pe} and PI both as nouns and adjectives.}.

Let $\sigma \in \mathcal{S}_n$ denote a permutation of $n$ elements. We define the action of $\sigma$ on an input $x = (x_1, x_2, \dots, x_n) \in X^n$ as
\(\sigma \cdot x = (x_{\sigma^{-1}(1)}, x_{\sigma^{-1}(2)}, \dots, x_{\sigma^{-1}(n)})\),
which permutes the components of $x$ according to $\sigma$.

\begin{definition}[\gls{pe} and \gls{pi}]\label{def:pe}
A function $f: X^n \to Y^n$ is said to be \gls{pe} if, for all $\sigma \in \mathcal{S}_n$,
\(
f(\sigma \cdot x) = \sigma \cdot f(x)
\).
A function $f: X^n \to Y$ is said to be \gls{pi} if, for all $\sigma \in \mathcal{S}_n$,
\(
f(\sigma \cdot x) = f(x)
\).
\end{definition}
In other words, \gls{pe} means that permuting the inputs causes the outputs to be permuted in the same way, while \gls{pi} means that permuting the inputs does not change the output at all.

Importantly, the composition of \gls{pe} functions is itself \gls{pe}:
\begin{proposition}\label{prop:compopefunc}
    If $f : X^n \to Y^n$, $g : Y^n \to Z^n$, and $h : X^n \to Y^n$ are \gls{pe}, so are $g \circ f$ and $f + h$.
    
\end{proposition}

\section{Related Work}

We briefly summarize related work on \gls{ctde} and neural networks that are \gls{pe}.

\paragraph{Progress on CTDE.}
In recent years, the \gls{ctde} paradigm gained significant attention.
Many studies have extended classical \gls{ctde} methods by integrating additional components.
For example, EMC~\cite{art:emc} introduces curiosity-driven exploration by using Q-value prediction error as an intrinsic reward; MASER~\cite{art:maser} generates subgoals from experience; and EOI~\cite{art:eoi} assigns agents to different roles.

Limitations of \gls{ctde} have been observed, though. Addressing the lack of policy diversity caused by parameter sharing, CDS~\cite{art:cds} incorporates agent-specific modules into a shared network. To mitigate the limits of decentralized execution, COLA~\cite{art:cola} proposes decentralized consensus, while CoCOM~\cite{art:cocom} builds on this idea with consensus-based communication.

However, most of these efforts remain within the \gls{ctde} framework.
In our work, we propose to take a step towards centralized execution, exploring the potential benefits of fully centralized policies.

\paragraph{Permutation Equivariant Neural Networks.}
\gls{pe} and \gls{pi} commonly arise in models for sets or graphs, where input order is irrelevant. DeepSets~\cite{art:deepset} introduced a general \gls{pi} framework using shared encoders followed by aggregation and transformation, and proposed a \gls{pe} variant using pooling within layers. GNNs and GCNNs~\cite{art:gnn,art:gcnn,art:gcnn2} are also inherently \gls{pe}/\gls{pi}, as their computations rely on graph connectivity rather than input order.

Other approaches include pairwise \gls{pe} networks~\cite{art:pairpe} that aggregate over element pairs, and modular constructions~\cite{art:pemab} combining shared encoders, pooling, and attention. In MARL, Hyper-Policy Networks~\cite{art:hpn} achieve \gls{pe} via hyper-networks generating agent-specific weights, with \gls{pi} obtained by summation.

\section{CPE Learning}
\label{sec:CPE}

In order to address the structural limitations of distributed policies, we introduce \emph{\gls{cpe}}, a novel MARL paradigm equipped with a centralized policy that follows \gls{cte} framework while incorporating the advantages of \gls{ctde}.

One of the main concerns with centralized policies is scalability.
% How distributed policy ensured scalability
A key scalability advantage of distributed policies in \gls{ctde} lies in their agent-number-agnostic design: with a fixed local observation space $Z_i$, the shared policy network $f: Z_i \to A_i$ can process $N$ observations to produce $N$ outputs for any number of agents $N$. 
The agent-number-agnostic nature of distributed policies ensures that the policy network structure does not directly depend on $N$, which significantly enhances scalability.

% Centralized policy lack...
By contrast, traditional centralized policies using classic networks such as \gls{mlp}s or RNNs concatenate agents’ observations into a large vector of size $N \cdot |z_i|$, with outputs of size $N \cdot |A_i|$. As $N$ increases, network size and parameter count grow rapidly, hindering training and generalization. Additionally, this approach ignores inherent structural symmetries in observations (e.g., the first two elements representing an agent's position), requiring the network to relearn these patterns from scratch.

\paragraph{Our Approach: CPE with GLPE Networks.}
We propose a centralized policy network that is both agent-number-agnostic and structure-preserving. Permutation Equivariant (PE) networks are a natural fit, as their design inherently maintains input symmetry and enables generalization to variable numbers of agents.

We aim to construct an agent-agnostic permutation equivariant (PE) structure. Let $D_n(X) \subseteq X^n \times X$ be defined as $\{((x_1, \ldots, x_n), x_i) \mid (x_1, \ldots, x_n) \in X^n,~ i \in \{1,\ldots,n\}\}$. Any PE network $f: X^n \to Y^n$ is entirely determined by
the function $g: D_n(X) \to Y$ defined by
\begin{equation}\label{eq:pedecomp}
   g(x, x_i) := f(x)_i\;.
\end{equation}
Note that $g$ is well-defined due to the fact that $f$ is \gls{pe}:
Let $\sigma_{ij}$ be the permutation that swaps the $i$-th and the $j$-th element. Then,
$x_i = x_j$ implies
$f(x)_{i}
= f(\sigma_{ij} \cdot x)_i
= (\sigma_{ij} \cdot f(x))_i
= f(x)_{j}$.
That is, the definition of $g$ does not depend on the index $i$.
Moreover, as a consequence of~\Cref{def:pe}, for a fixed $x_i$, the function $g(\,\cdot\, ,x_i)$ is permutation-invariant (PI) in $x$:
$g(\sigma \cdot x, x_i) = g(x, x_i)$.

From~\cite{art:deepset}, for a fixed $n$, any PI function $h: X^n \to Y$ can be approximated arbitrarily close by a function of the form:
\begin{equation}\label{eq:piapprox}
    h(x) = \rho\left(\sum_{j=1}^n \phi(x_j)\right),
\end{equation}
for suitable transformations $\rho$ and $\phi$. 
Although $h$ is defined on $X^n$, we can define the \gls{pi} function $h': X^{n'} \to Y$ with $n' \neq n$ as the natural extension of $h$ using the same $\rho$ and $\phi$ function via $h'(x') = \rho\left(\sum_{j=1}^{n'} \phi(x'_j)\right).$
This means that a \gls{pi} function can be approximated by an $n$-agnostic network structure composed of networks presenting $\rho$ and $\phi$.

For now, we are not making any assumption on the relation between $h$ and $h'$, but aim only to show that a \gls{pi} function $h$ can be approximated in an $n$-agnostic way, i.e., its network structure does not depend on $n$.\footnote{It is an interesting direction for future work to study the relation between $h$ and $h'$ when using the same trained network $\rho$ and $\phi$, especially in the context of transfer learning.}

Extending \eqref{eq:piapprox} to $g(x, x_i)$, since $g(x,x_i)$ is PI with respect to $x$ when $x_i$ is fixed, we write:
\[
g(x, x_i) = \rho_{x_i}\left(\sum_{j=1}^n \phi_{x_i}(x_j)\right).
\]
This universal representation requires defining $\rho_{x_i}$ and $\phi_{x_i}$ for each $x_i$, which is infeasible when $X$ is uncountable. 
On the other hand, using fixed $\rho$ and $\phi$ independent of $x_i$ reduces $g$, and consequently $f$, to a purely \gls{pi} function. 
To maintain $n$-agnosticism while allowing $\rho_{x_i}$ and $\phi_{x_i}$ to vary with $x_i$, we adopt a computationally efficient approach: we share the same $\rho$ and $\phi$ networks across all $x_i$, and introduce additive biases $b_{\rho}$ and $b_{\phi}$ conditioned on $x_i$:
\begin{equation}\label{eq:peagno}
    g(x, x_i) = \rho\left(\sum_{j=1}^n \left( \phi(x_j) + b_\phi(x_i) \right)\right) + b_\rho(x_i)\;.
\end{equation}

Thus, we obtain a simple framework for implementing an $n$-agnostic PE function. From \eqref{eq:peagno}, we identify two main components in this design: 
$\sublayer_{\text{glo}}(x, x_i) = \rho\left(\sum_{j=1}^n \left(\phi(x_j) + b_{\phi}(x_i) \right)\right)$ and $\sublayer_{\text{loc}}(x_i) = b_{\rho}(x_i)$.
Here, $\sublayer_{\text{loc}}$ captures purely local information, while $\sublayer_{\text{glo}}$ aggregates information from the joint input.

Inspired by this structure and \cite{art:deepset}, we propose the \gls{glpe} network, an agent-number-agnostic architecture for centralized policy implementation, composed of \gls{glpe} layers. Each layer $f_{\text{GLPE}}$ follows the $n$-agnostic design with two components:
\begin{itemize}
\item[(i)] A local sub-layer $\sublayer_{\text{loc}}: X \to Y$ that captures individual features. We model this as a one-layer neural network: $\sublayer_{\text{loc}}(x_i) = \sublayerfunc(x_i)$, corresponding to $b_\rho(x_i)$ in \eqref{eq:peagno}.

\item[(ii)] A global sub-layer $\sublayer_{\text{glo}}: X^n \to Y$ that aggregates joint features. Instead of sum-pooling, we apply mean-pooling, which preserves PI and $n$-agnosticity, while improving generalization across different $n$. Noting that
$
\frac{1}{n}\sum_{j=1}^n \left(\phi(x_j) + b_{\phi}(x_i) \right) = \frac{1}{n}\sum_{j=1}^n \phi(x_j) + b_{\phi}(x_i),
$
we separate $b_{\phi}(x_i)$ from $\rho$ and merge it into the local term. We then define $\sublayer_{\text{glo}}(x) = \tanh\left(\sublayerfunc_{\text{pooling}}(\text{mean}(x))\right)$, where $\sublayerfunc_{\text{pooling}}$ is a one-layer network.
\end{itemize}
The resulting \gls{glpe} layer is:
\begin{align*}
f_{\text{GLPE}}(x) 
&= \left\{ \sublayer_{\text{loc}}(x_i) + \sublayer_{\text{glo}}(x) \right\}_i \nonumber \\
&= \left\{ \sublayerfunc(x_i) + \tanh\left(\sublayerfunc_{\text{pooling}}(\text{mean}(x)) \right) \right\}_i\;.
\end{align*}

\begin{figure*}[t]
    \centering
    \includegraphics[width=0.9\linewidth]{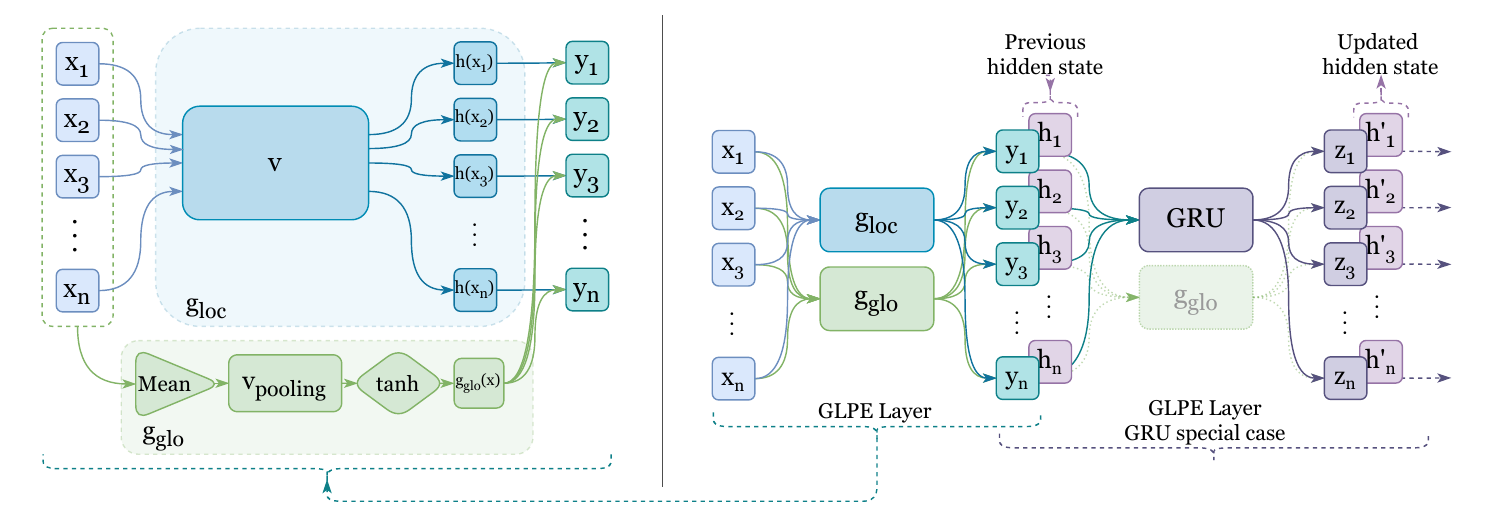}
    \caption{Illustration of the \gls{glpe} network structure.  
    Left: Zoomed-in view of the internal structure of a general \gls{glpe} layer.
    Right: Two stacked \gls{glpe} layers. A general layer ($1^\text{st}$ layer) and a GRU-based special case.}
    \label{fig:glpe}
\end{figure*}

Since both sub-layers are shared across agents, the \gls{glpe} structure scales well and can be illustrated as in~\Cref{fig:glpe}.

We made several specific design choices for the \gls{glpe} layers. First, each neural network component---$\sublayerfunc$ and $\sublayerfunc_{\text{pooling}}$---is set to a single-layer \gls{mlp}. While deeper architectures can also maintain \gls{pe}, shallow networks enable earlier injection of global context into local features. This design better reflects the ``layer'' abstraction and simplifies integration with other modules, such as RNNs, in subsequent layers.
Second, we replace the sum pooling in the global sub-layer protocol with mean pooling. While sum, mean, or max pooling all preserve \gls{pi} and agent-number-agnostic properties, mean pooling avoids the differentiability issue of max and the scaling issue of sum, where outputs grow with the number of agents.
Third, we apply a $\tanh$ activation to the global sub-layer’s output to bound its values. This encourages the model to focus more on local observations, which are often more informative for agent-specific decisions in MARL. We empirically validate the choice of mean pooling and $\tanh$ in a toy example (see Supplementary Material).

It's worth mentioning that the \gls{glpe} structure is not a universal approximator of permutation equivariant functions, especially after introducing $\tanh$, but rather a specialized instantiation of \eqref{eq:peagno} tailored for MARL.
It captures global context from joint observations, a key advantage of centralized policies, while preserving the structure of individual agent inputs. It remains lightweight and scalable due to its agent-number-agnostic design and can be easily implemented. 
% It also integrates well with specialized \gls{ctde} architectures.

By \Cref{prop:compopefunc}, \gls{glpe} layers can be stacked to form deeper \gls{pe} networks. In \gls{cpe}, we replace traditional distributed policy networks $f_{\text{Distributed}}: Z_i \to A_i$ in \gls{ctde} algorithms (e.g., distributed Q-networks or actors) with \gls{glpe} networks $f_{\text{GLPE}}: Z_i^N \to A_i^N$. When a \gls{ctde} policy includes modules such as GRUs, we incorporate them directly into the \gls{glpe} structure by treating them as local-only layers without a global sub-layer: $\sublayer_{\text{loc}}(x_i, h_i) = \text{GRU}(x_i, h_i)$ and $\sublayer_{\text{glo}}(x)=0$. \Cref{fig:glpe} illustrates an example of this integration.

CPE retains the centralized training components of \gls{ctde} methods. While these components were originally designed to coordinate distributed agents under partial observability during training, they offer benefits beyond that. For example, the QMIX mixer decomposes the global reward into agent-specific contributions, providing more informative and fine-grained credit assignment than naïve global targets. Moreover, centralized components often leverage additional information from the global state, which can also benefit centralized \gls{glpe} policies that only access joint observations during execution. These richer training signals enhance policy learning. Therefore, we adopt the training mechanisms of \gls{ctde} algorithms and apply them to our centralized \gls{glpe} policies.

% homogeneity among agents: otherwise HAPPO is also \gls{ctde}
Importantly, CPE is compatible with any \gls{ctde} algorithm that updates policies in a centralized and synchronized manner. A counterexample is, HAPPO~\cite{art:happo} performs sequential individual policy updates, making it incompatible with the current CPE framework based on \gls{glpe} networks. For methods that incorporate custom components into policy networks, CPE can be adapted accordingly, as demonstrated in our GRU integration.

\section{Experimental Evaluation}
We evaluated the CPE approach in three challenging MARL environments: \gls{mpe}, \gls{smac}, and \gls{rware}.
To show that our approach is feasible, we first compared the \gls{glpe} network to a \gls{mlp} centralized policy. 
Then, we compared CPE-adapted algorithms to their \gls{ctde} counterpart on all games. The code is available in the supplementary material.

\paragraph{MARL environments.}
We evaluate CPE on three diverse benchmarks that pose distinct coordination and observability challenges. \emph{Spread} from \gls{mpe} involves $n$ agents covering $n$ landmarks without collisions; we disable inter-agent observations to simulate partial observability. \emph{SMAC}~\cite{art:smac} is a StarCraft II-based environment featuring high-dimensional inputs and strategic combat coordination. \emph{RWARE} simulates a warehouse with sparse rewards, large observation spaces, and complex agent interactions. Together, these tasks span a broad range of difficulty and structural properties, making them well-suited to assess the flexibility and robustness of the CPE framework.

\paragraph{Benchmark setup.}
We selected two popular \gls{ctde} algorithms from the value decomposition family (QMIX and QLPEX) and two actor-critic methods (MAA2C and MAPPO), and applied the CPE framework to them. Following the original designs, the CPE versions of the Q-network and actor consist of three layers, including a GRU unit, all built using \gls{glpe} structures. To ensure fair comparison, we match the hidden dimensions to those of the original networks.

Each algorithm is trained for $10.05$M steps on \gls{smac}, and $40.05$M steps on \gls{mpe} and \gls{rware}, using $5$ random seeds. Every $10$k steps in \gls{smac} and $50$k in \gls{mpe}/\gls{rware}, we conduct evaluation using $32$ and $100$ test episodes respectively.
% recording cumulative episodic rewards.
Our main metric is the mean test win rate (\gls{smac}) or mean test episodic reward (\gls{mpe}/\gls{rware}), averaged over seeds. We report the mean and $75\%$ confidence interval. 
Hyperparameters follow Pymarl3~\cite{art:hpn} for \gls{smac} and Pymarlzooplus~\cite{art:pymarlzooplus} for \gls{mpe}/\gls{rware}. CPE uses identical configurations to ensure fair comparison. Full details and best-policy results, defined by highest average reward over the final $10$ test steps, are in the appendix.

% \gls{mlp} vs \gls{glpe}
\paragraph{Ablation Study: \gls{mlp} vs.\ \gls{glpe}.}
Before comparing \gls{ctde} methods with those using CPE, we first validated our choice of \gls{glpe} as the centralized policy network.  
We implemented two versions of QMIX with centralized policies: one using a \gls{glpe} network, and the other using a standard \gls{mlp}. Both models are feedforward (i.e., no recurrent units) and consist of three layers with the same hidden dimension. In the \gls{mlp} policy, inputs and outputs are 1D tensors, reshaped appropriately before and after processing. These two variants are evaluated on the \gls{mpe} Spread task.

\begin{figure}[t]
    \centering
    \includegraphics[width=0.9\linewidth]{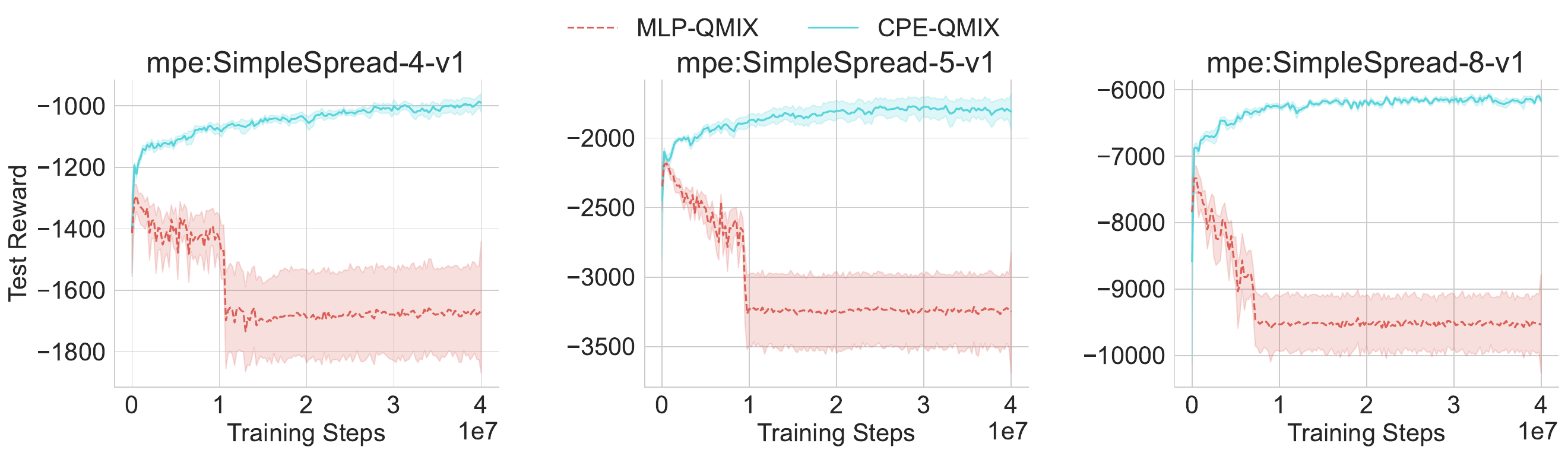}
    \caption{Test-time reward of \gls{mlp} and \gls{glpe} centralized policies on Spread-\{4,5,8\}.}
    \label{fig:centspread}
\end{figure}

We compared the reward and the number of parameters of the centralized Q-networks in CPE-QMIX and MLP-QMIX, as shown in \Cref{fig:centspread} and \Cref{tab:centpolicyparam}. As expected, CPE-QMIX achieves constantly higher performance, and despite both architectures using the same hidden dimensions, the parameter count in the \gls{mlp} policy increases significantly faster with the number of agents compared to the \gls{glpe}-based policy.

\begin{table}[hbt]
    \centering
    \small
    \setlength{\tabcolsep}{1mm}
    \begin{tabular}{lcc}
        \toprule
        \textbf{Scenario} & \textbf{MLP Policy} & \textbf{GLPE Policy} \\
        \midrule
        Spread-4 & 10.90K & 11.65K \\
        Spread-5 & 13.53K & 12.04K \\
        Spread-8 & 23.72K & 13.19K \\
        \bottomrule
    \end{tabular}
    \caption{Number of parameters in centralized policy networks for \gls{mlp} and \gls{glpe} across \gls{mpe} Spread scenarios.}
    \label{tab:centpolicyparam}
\end{table}

% \gls{smac}
\paragraph{\gls{smac}.}
We selected four super-hard maps, each posing different challenges: \texttt{6h\_vs\_8z} as a classic battle scenario, \texttt{3s5z\_vs\_3s6z} to test whether CPE can handle heterogeneous units, \texttt{27m\_vs\_30m} to assess scalability to environments with more agents, and \texttt{corridor} to evaluate coordination capabilities in a narrow map.

\begin{figure}[t]
    \centering
    \includegraphics[width=0.9\linewidth]{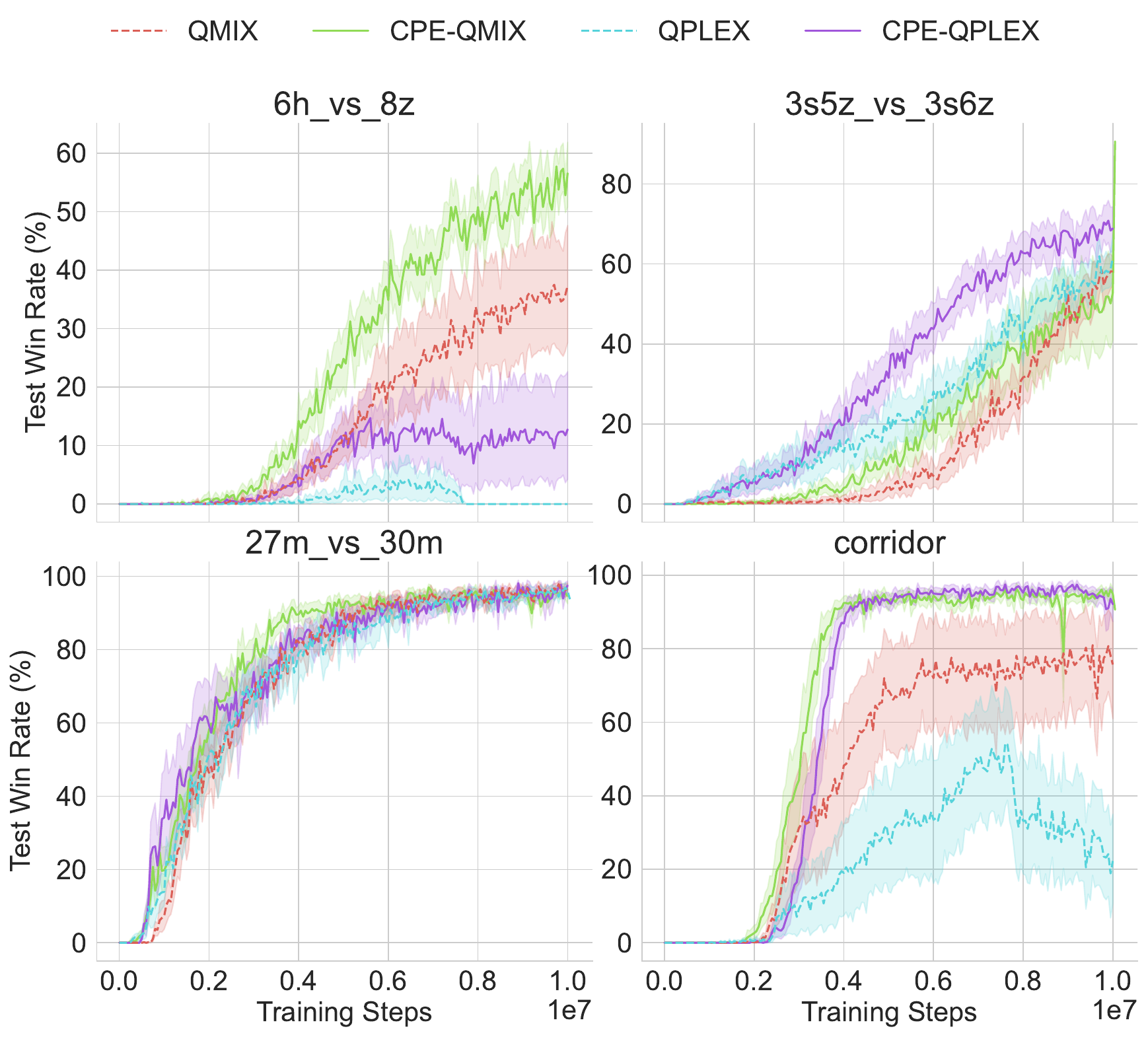}
    \caption{Mean win rate of CPE and vanilla methods on different super hard \gls{smac} maps.}
    \label{fig:smac}
\end{figure}

We tested two variants of our method, CPE-QMIX and CPE-QPLEX, against their vanilla counterparts (\Cref{fig:smac}). 
Across all selected maps, CPE methods demonstrated consistent performance improvements, showing their adaptability to various complex tasks. On \texttt{27m\_vs\_30m}, both CPE methods achieve the same converged win rate of $1.0$ as their baselines, and exhibit faster learning in early training stages, suggesting that CPE not only scales well with more agents but also enhances sample efficiency. The strong performance on \texttt{3s5z\_vs\_3s6z} indicates that the \gls{glpe} policy network can generalize to scenarios with heterogeneous units, as long as their observation and action spaces are aligned. In \texttt{corridor}, where effective coordination is critical due to the narrow terrain and high agent density (20 units per side), CPE significantly outperforms the baselines. This highlights the strength of centralized policies in handling coordination, especially in crowded and high-interaction environments.

The number of parameters processed by distributed policies (QMIX and QPLEX) and the \gls{glpe} policies are shown in~\Cref{tab:smac_param}. As the number of agents increases across different maps, both types of policies see a rise in parameter count, primarily because the size of individual observation spaces in \gls{smac} scales with the number of agents. Although \gls{glpe} policies grow faster than their distributed counterparts, the increase remains relatively modest. Empirically, the increase in parameter count for \gls{glpe} policies is consistently around twice that of distributed policies when comparing different maps. 
Given the same hidden dimensions and number of layers, we formally show in the appendix that \gls{glpe} networks have at most twice as many parameters as distributed policies: $|\theta_{\text{GLPE}}| \leq 2 \cdot |\theta_{\text{Distributed}}|$.

\begin{table}[t]
    \centering
    \small
    \setlength{\tabcolsep}{1mm}
    \begin{tabular}{lcc}
        \toprule
        \textbf{Map} & \textbf{Distributed Policy} & \textbf{GLPE Policy} \\
        \midrule
        6h\_vs\_8z & 31.31K & 37.58K \\
        3s5z\_vs\_3s6z & 35.22K & 45.39K \\
        27m\_vs\_30m & 47.33K & 69.60K \\
        corridor & 37.34K & 49.63K \\
        \bottomrule
    \end{tabular}
    \caption{Number of parameters for distributed and \gls{glpe} policies across different \gls{smac} maps.}
    \label{tab:smac_param}
\end{table}

\begin{figure*}[t]
    \centering
    \includegraphics[width=0.98\linewidth]{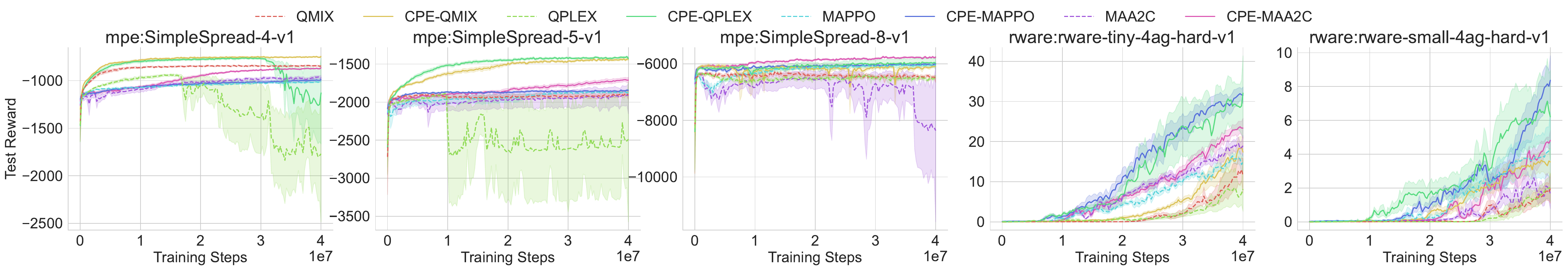}
    \caption{Results on \gls{mpe} Spread with varying numbers of agents (left three plots), and on \gls{rware} with two different map sizes and 4 agents (right two plots).}
    \label{fig:mperware}
\end{figure*}

% CPE vs \gls{ctde}
\paragraph{\gls{mpe}.}
We evaluate the algorithms with varying numbers of agents, $N \in \{4, 5, 8\}$, to reflect different levels of task difficulty and assess the scalability of the CPE framework. The mean rewards are shown in \Cref{fig:mperware}. In all scenarios, CPE-enhanced algorithms consistently outperformed their vanilla counterparts, achieving the highest overall rewards. Notably, substantial performance gains were observed for QMIX, QPLEX, and MAA2C, demonstrating that CPE benefits both value decomposition and actor-critic methods. Moreover, some baseline \gls{ctde} algorithms, such as QPLEX and MAA2C, exhibited unstable learning dynamics with large reward variance. The centralized policy introduced by CPE helps stabilize learning, potentially by mitigating the effects of non-stationary policies of other agents, which would otherwise be treated as part of the evolving environment.

In terms of best final rewards, CPE yielded maximum improvements of $168.76$ (QPLEX), $327.34$ (QMIX), and $958.24$ (QPLEX) on Spread-4, Spread-5, and Spread-8, respectively. These results suggest that the \gls{glpe}-based centralized policy not only maintains its advantage but strengthens it as the number of agents increases, likely due to the increasing complexity of coordination required in larger populations. 

Furthermore, as shown in~\Cref{tab:gympolicyparam}, the agent-number agnostic nature of GLPE ensures that doubling the number of agents from 4 to 8 increases the number of policy parameters by only 5.73\%, primarily due to the enlarged observation space. 
% (e.g., additional destinations). 
This highlights the scalability of the \gls{glpe}-based centralized policy.

\paragraph{\gls{rware}.}  
We evaluated two scenarios with four agents on the ``tiny'' and ``small'' maps, both under the most challenging ``hard'' setting. In both cases, integrating CPE consistently led to performance gains for all baseline \gls{ctde} algorithms. Even in the least favorable instance, CPE maintained performance comparable to the original methods (see reward curves in \Cref{fig:mperware}).

On the tiny map, CPE improved QPLEX and MAPPO's performance by over $200\%$, surpassing the previously reported SOTA score of $27.85 \pm 19.71$ achieved by MAT-DEC~\cite{art:matdec,art:pymarlzooplus}. These results confirm that CPE effectively enhances a wide range of baseline methods, both value-decomposition and actor–critic methods, delivering substantial improvements across different scenarios. 

% Number of parameters
Parameter counts are reported in~\Cref{tab:gympolicyparam}. To accommodate the larger observation spaces in \gls{rware} compared to \gls{mpe}, we doubled the hidden dimensions of all policy networks for this environment. Even with this adjustment, the size difference between the distributed policies and the \gls{glpe}-based centralized policy remains moderate, at most $12.82\%$, demonstrating the adaptability of \gls{glpe} to varying input dimensionalities.

\begin{table}[t]
    \centering
    \small
    \setlength{\tabcolsep}{1mm}
    \begin{tabular}{lcc}
        \toprule
        \textbf{Scenario} & \textbf{Distrib.\ Policy} & \textbf{\gls{glpe} Policy} \\
        \midrule
        Spread-4 & 26.693K & 28.357K \\
        Spread-5 & 26.885K & 28.741K \\
        Spread-8 & 27.461K & 29.893K \\
        rware-tiny-4ag-hard-v1 & 112.645K & 126.085K \\
        rware-small-4ag-hard-v1 & 113.797K & 128.389K \\
        \bottomrule
    \end{tabular}
    \caption{Number of parameters in centralized policy networks for distributed and \gls{glpe} across different scenarios.}
    \label{tab:gympolicyparam}
\end{table}

Overall, the results demonstrate that CPE is a general-purpose enhancement module that can be seamlessly integrated into existing \gls{ctde} algorithms.

\paragraph{Policy Quality.}
We inspected the policies learned with and without CPE. We selected the best-trained models of QPLEX and CPE-QPLEX on \texttt{rware-tiny-4ag-hard-v1}, which achieved average test rewards of $23.20$ and $44.57$, respectively. We ran $10$ evaluation episodes and collected each agent’s observations along with the selected actions.

\begin{figure}
    \centering
    \includegraphics[width=0.9\linewidth]{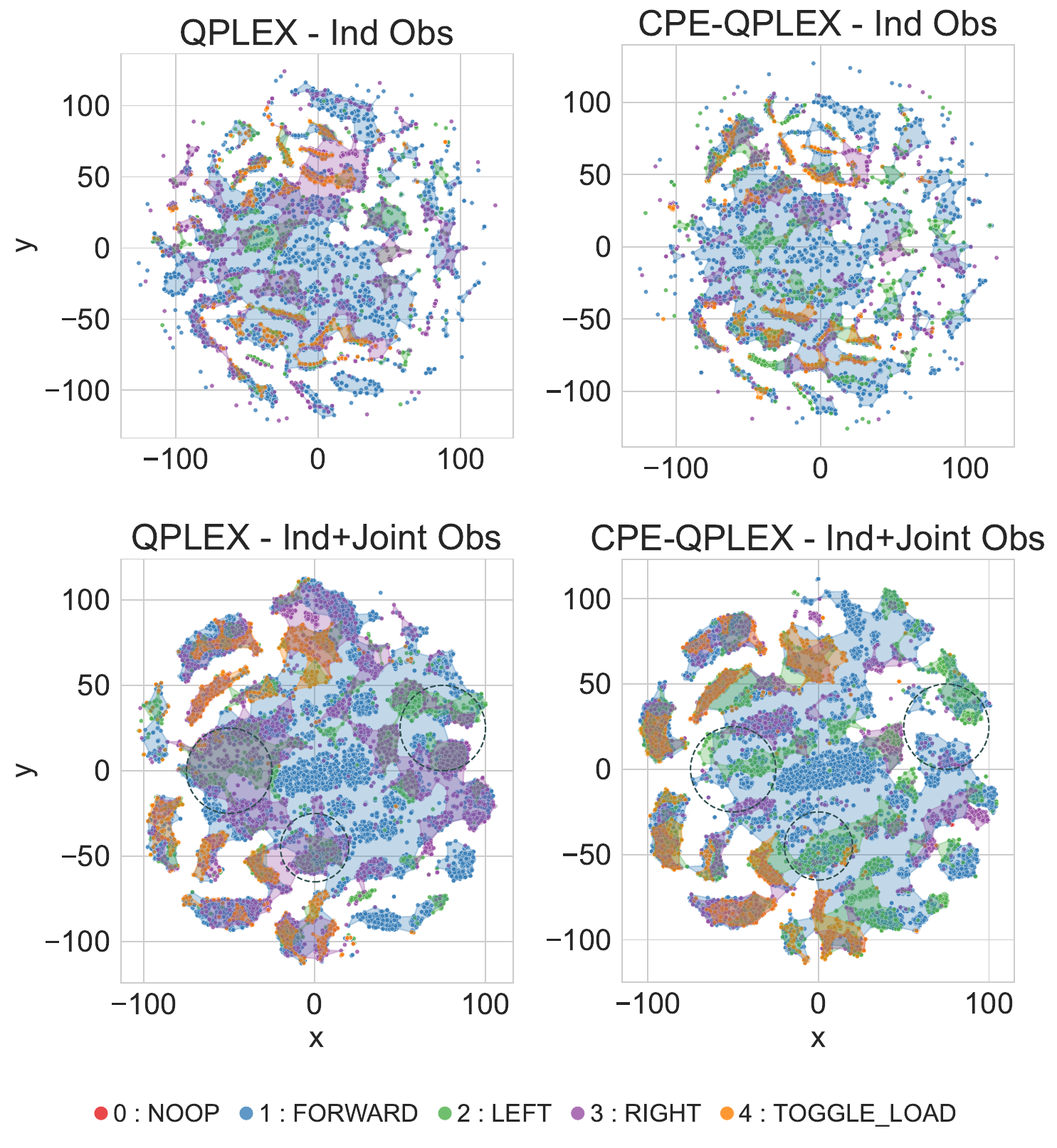}
    \caption{2D t-SNE embeddings of individual observations (top) and the concatenation of individual observations with their global mean (bottom) for QPLEX and CPE-QPLEX.}
    \label{fig:snepolicy}
\end{figure}

To visualize the agents' policies, we first embedded the high-dimensional observation vectors into two dimensions using t-SNE~\cite{art:tsne}, and colored each point according to the agent's selected action.  For each action category, we applied DBSCAN~\cite{art:dbscan} to identify dense clusters. Finally, we used the alpha-shape algorithm to draw boundaries around these clusters, providing a clearer spatial interpretation of the agent behavior. The results are presented in~\Cref{fig:snepolicy}. A higher-quality policy is expected to produce clusters with clearer boundaries and less overlap between opposing actions, particularly between ``turn left'' (green) and ``turn right'' (purple), which represent conflicting strategies in the \gls{rware} environment.

When using only individual observations as input, both policies produce similar clustering patterns. While the clusters for CPE-QPLEX appear slightly more compact, the overlap between green and purple remains limited for both policies. However, when we concatenate the mean of all individual observations, representing the joint observation, with each local observation and redo the embedding, the difference becomes more evident. In the QPLEX case, clusters expand more and the overlap between green and purple increases, particularly in the circled regions. 
This suggests that joint observations, unavailable to distributed policies, contain critical information for resolving action ambiguities that cannot be disambiguated from local inputs alone. The \gls{glpe}-based centralized policy leverages this global context to make more informed decisions.

\section{Conclusion}
We revisited the dominant \gls{ctde} paradigm in MARL and proposed Centralized Permutation Equivariant (CPE) learning, a new framework that overcomes the scalability limitations of traditional centralized policies. By exploiting symmetries inherent in multi-agent environments, CPE introduces an agent-number-agnostic architecture that improves the performance of standard \gls{ctde} methods, both value-based and actor-critic, without incurring scalability issues.
Although in this work we applied CPE to four widely used algorithms, the approach is broadly applicable and can be extended to other \gls{ctde} methods. As future work, we plan to investigate the effectiveness of CPE under curriculum learning setups, where generalization across varying numbers of agents is essential.

\section*{Acknowledgment}
This work was supported by the French National Research Agency (ANR) through the DREAMY project (ANR-21-CE48-0003) and the SAIF project, funded by the ``France~2030'' government investment plan and managed by ANR (ANR-23-PEIA-0006). It was carried out with the help of HPC resources provided by GENCI–IDRIS (Grant 2025-AD011014516R1).

% \clearpage 

\bibliography{aaai2026}

\clearpage 

\appendix
\section*{Appendix for \textit{``Centralized Yet Effective: Revisiting Policy Design for Cooperative Multi-Agent Learning''}}

\vspace{2em}

\section{Toy Test for \gls{glpe} Structure}

This toy example is designed to quickly validate our GLPE network structure, especially the choice of pooling function for $\sublayer_{\text{glo}}$ and the addition of $\tanh$ activation. The task involves joint inputs $x$ composed of $N$ individual inputs $x_i$ of length $d$, where each entry is randomly sampled between $0$ and $N$. Each $x_i$ has a corresponding output $y_i$ defined under different test modes:

\begin{itemize}
    \item Mean: $y_i = x_i + (\sum_{j=1}^N x_j)/N$
    \item Sum: $y_i = x_i + \sum_{j=1}^N x_j$
    \item Max: $y_i = x_i + \max_{j=1}^N x_j$
\end{itemize}

These settings mimic our assumptions in MARL, where the individual output depends both on its own input and joint input statistics. Notably, each $x_i$ is scaled with $N$ to prevent the joint terms from dominating the output. We test seven models with $3$ layers (ELU activations, hidden size $64$), including MLP and PE variants using mean, sum, or max pooling, with or without $\tanh$ in the global sublayer. We fix $d=2$ and vary $N \in \{5, 10, 20\}$ to test scalability.

\begin{figure*}[t]
    \centering

    % First row
    \begin{subfigure}[b]{0.32\textwidth}
        \includegraphics[width=\textwidth]{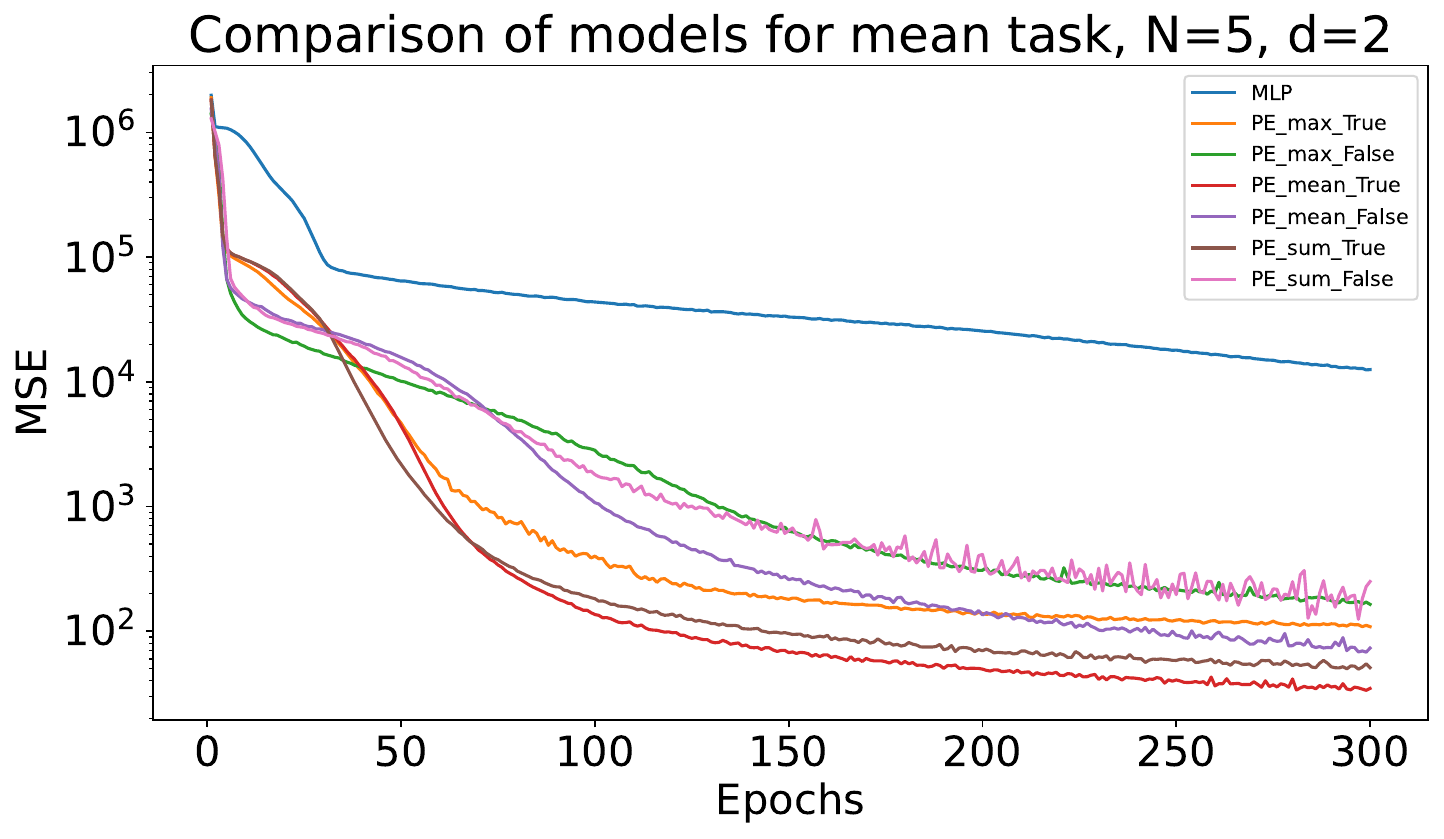}
        \caption{$N=5$}
    \end{subfigure}
    \hfill
    \begin{subfigure}[b]{0.32\textwidth}
        \includegraphics[width=\textwidth]{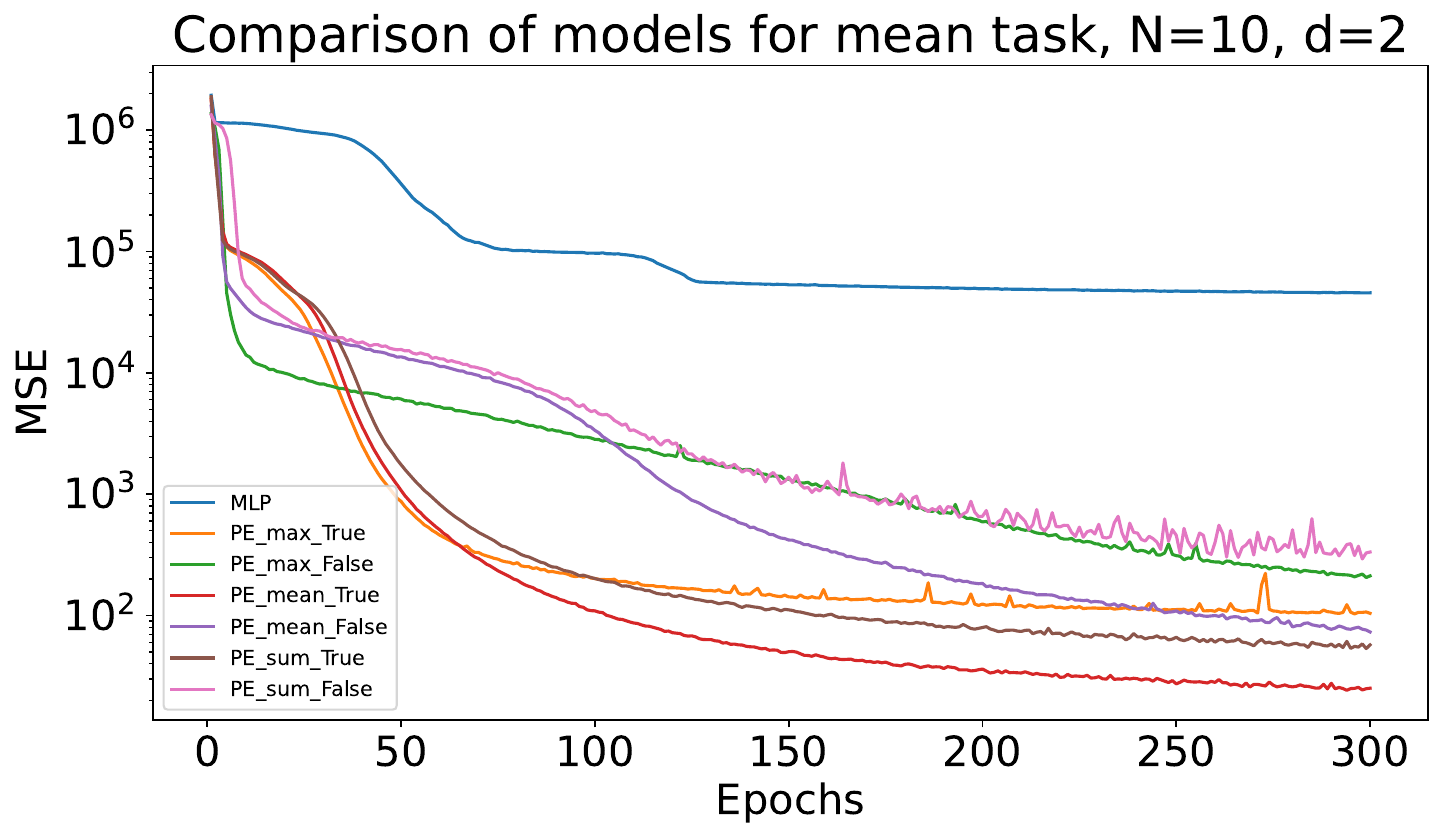}
        \caption{$N=10$}
    \end{subfigure}
    \hfill
    \begin{subfigure}[b]{0.32\textwidth}
        \includegraphics[width=\textwidth]{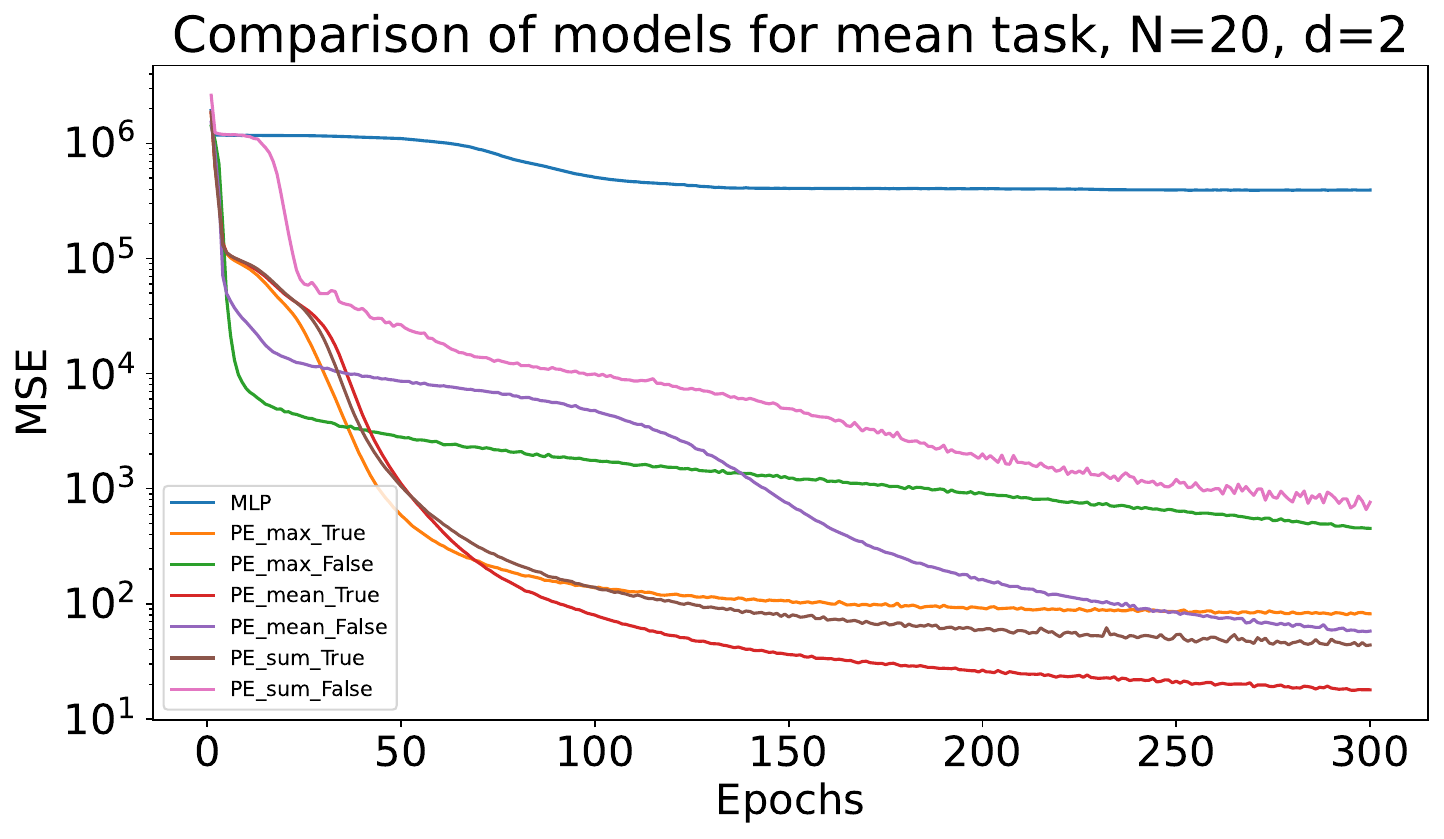}
        \caption{$N=20$}
    \end{subfigure}

    \vspace{1em}

    % Second row
    \begin{subfigure}[b]{0.32\textwidth}
        \includegraphics[width=\textwidth]{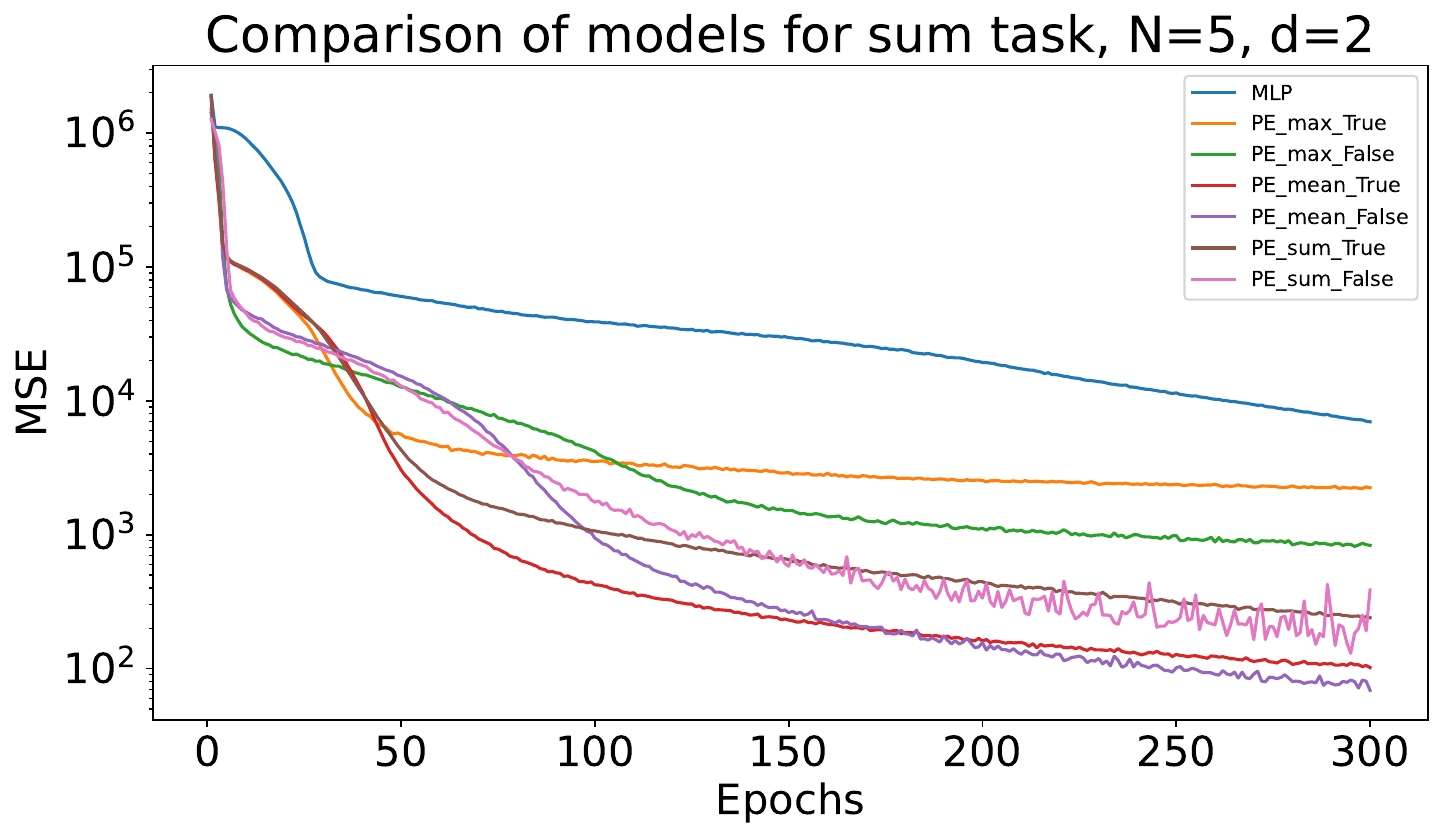}
        \caption{$N=5$}
    \end{subfigure}
    \hfill
    \begin{subfigure}[b]{0.32\textwidth}
        \includegraphics[width=\textwidth]{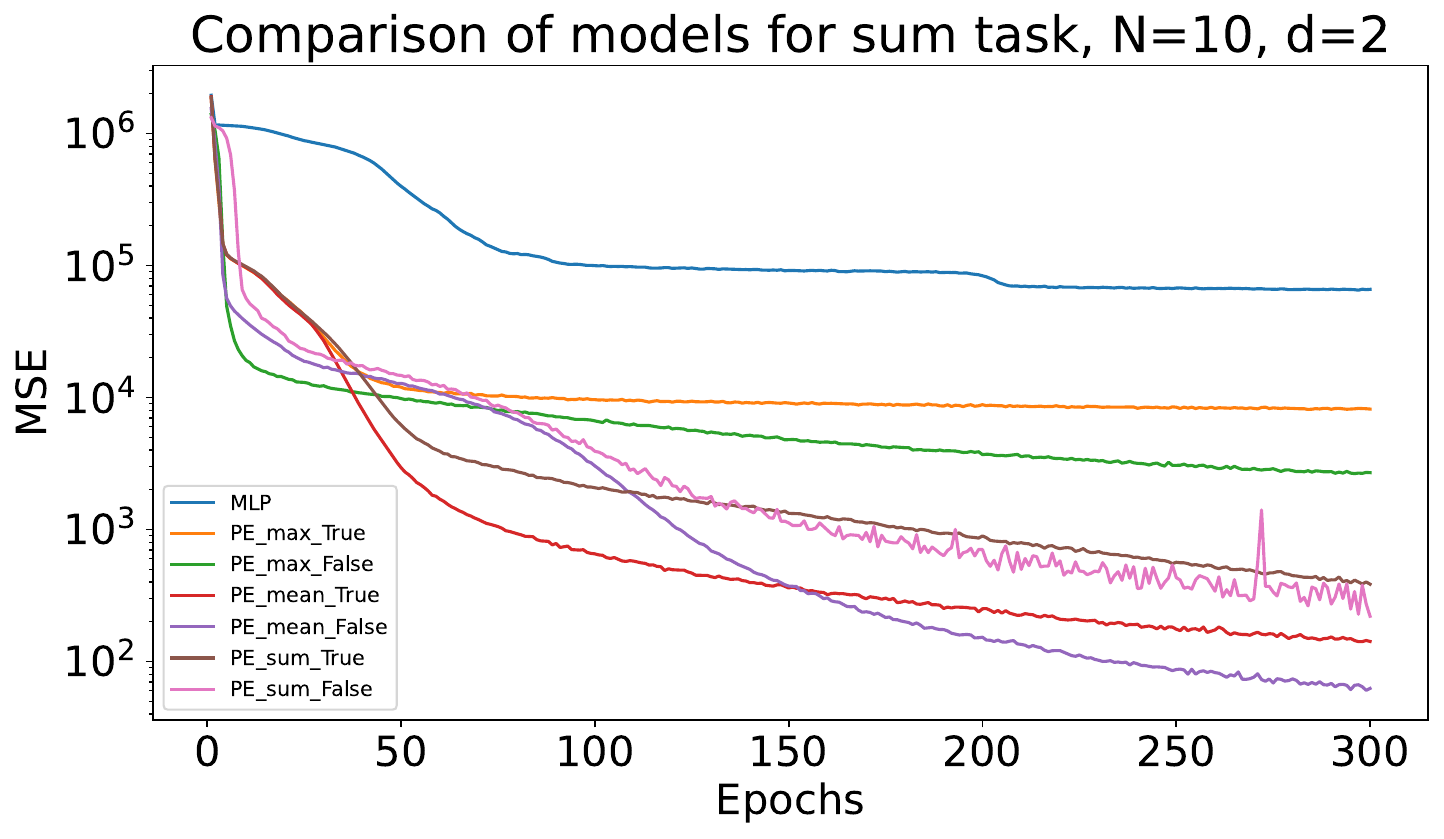}
        \caption{$N=10$}
    \end{subfigure}
    \hfill
    \begin{subfigure}[b]{0.32\textwidth}
        \includegraphics[width=\textwidth]{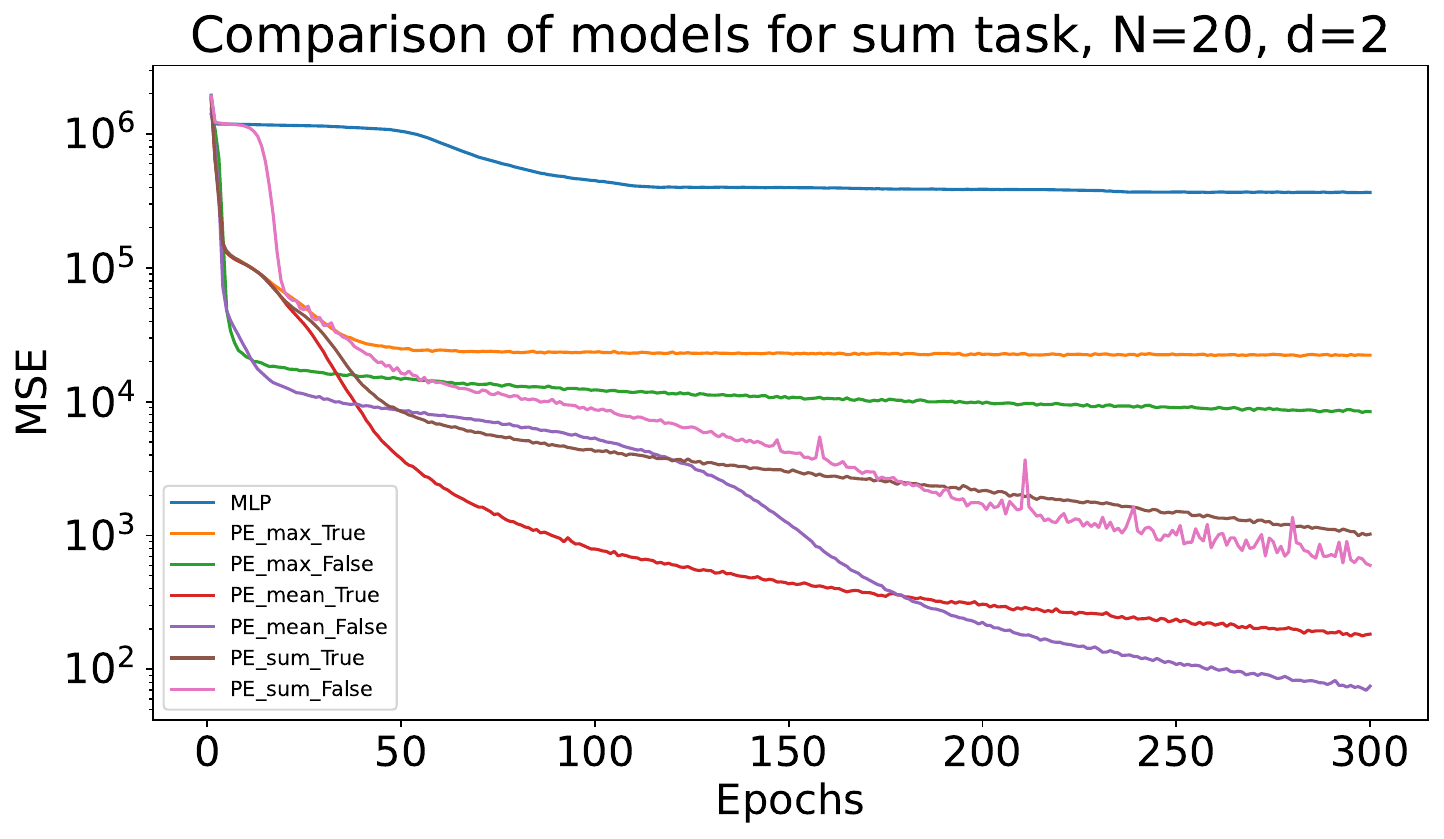}
        \caption{$N=20$}
    \end{subfigure}

    \vspace{1em}

    % Third row
    \begin{subfigure}[b]{0.32\textwidth}
        \includegraphics[width=\textwidth]{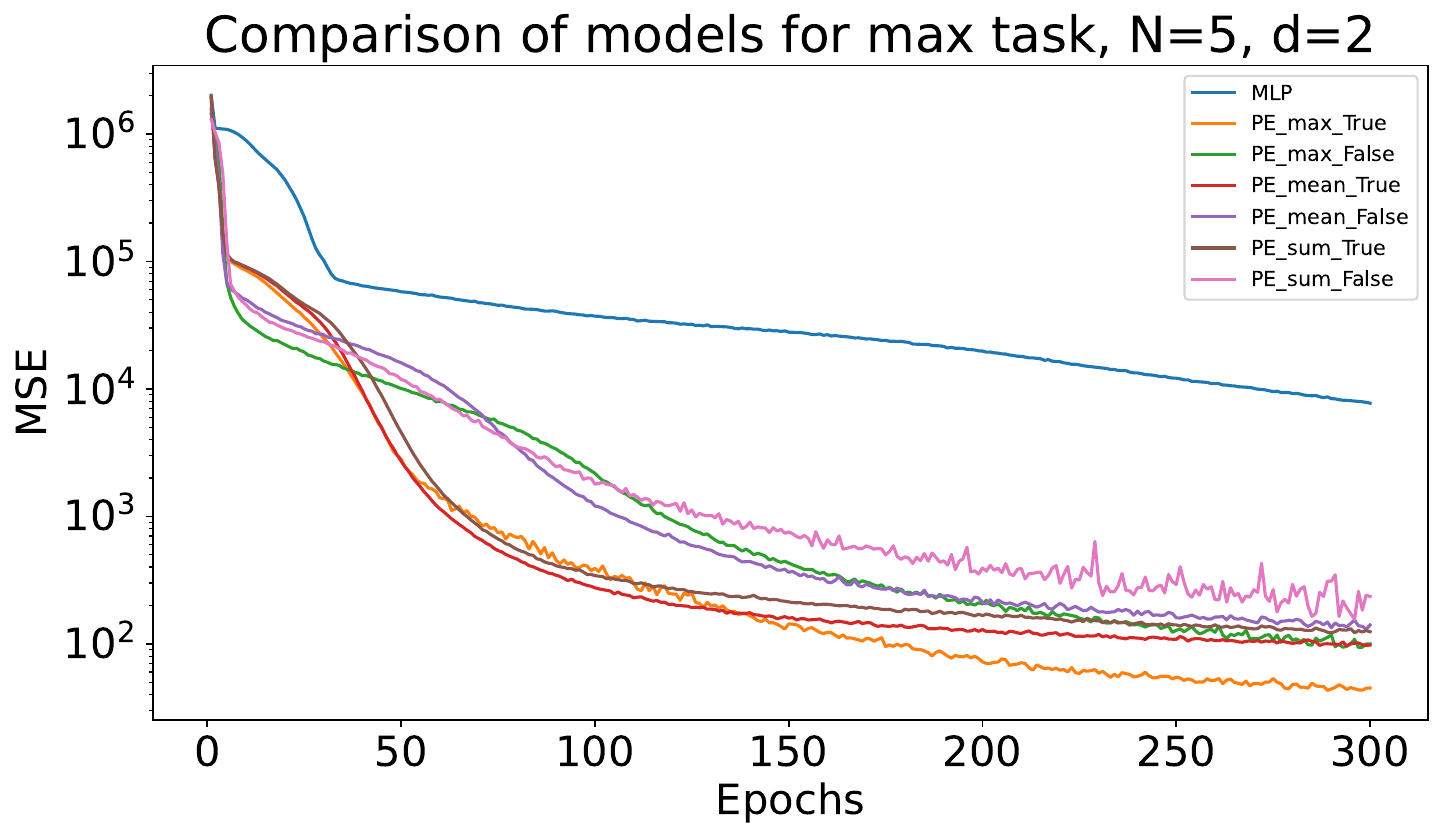}
        \caption{$N=5$}
    \end{subfigure}
    \hfill
    \begin{subfigure}[b]{0.32\textwidth}
        \includegraphics[width=\textwidth]{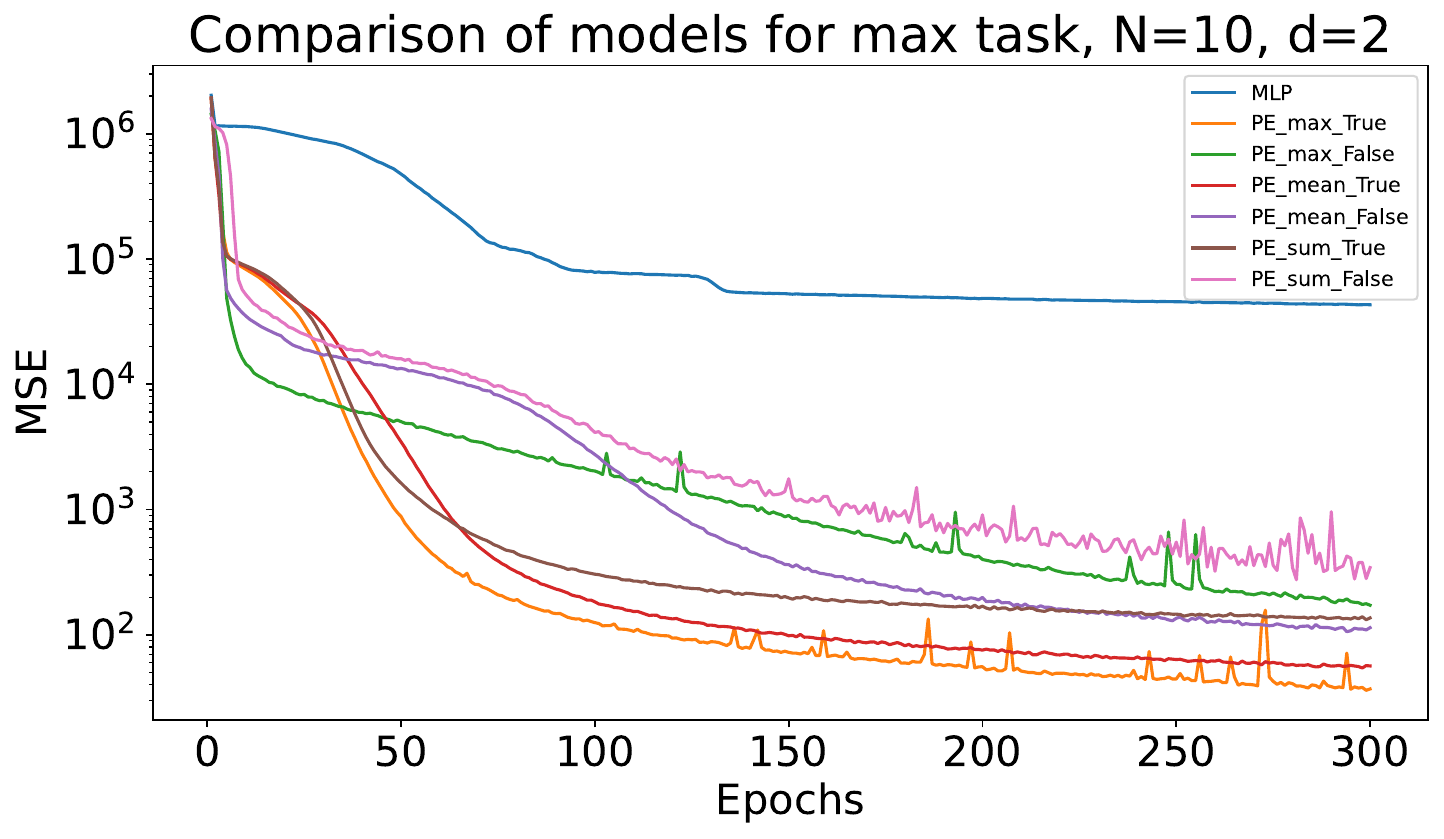}
        \caption{$N=10$}
    \end{subfigure}
    \hfill
    \begin{subfigure}[b]{0.32\textwidth}
        \includegraphics[width=\textwidth]{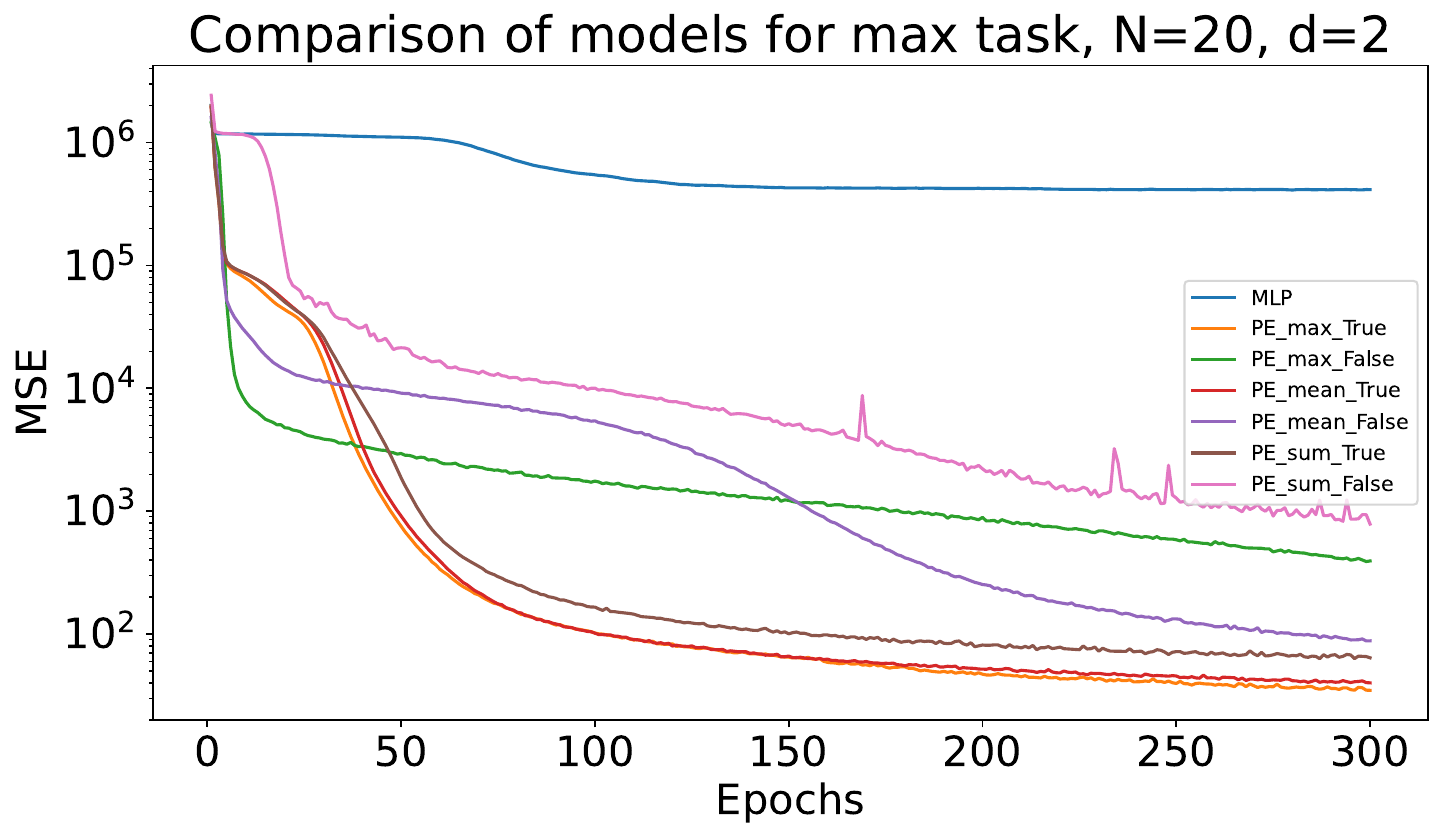}
        \caption{$N=20$}
    \end{subfigure}

    \caption{Experiments on three functions: mean (top), sum (middle), and max (bottom).}
    \label{fig:toy_all}
\end{figure*}

Each model is trained for $300$ epochs with $32$ joint inputs per epoch. Results (averaged over $3$ random seeds) are shown in \Cref{fig:toy_all}.

The best-performing models are: mean mode — PE with mean pooling and $\tanh$; sum mode — PE with mean pooling without $\tanh$; and max mode — PE with max pooling and $\tanh$. While we initially expected matching pooling types to perform best in corresponding tasks, mean pooling surprisingly outperforms sum pooling in the sum task. In the max task, although mean pooling does not reach the best loss, its performance improves with increasing $N$.

Overall, $\tanh$ activation enhances scalability: models with $\tanh$ outperform counterparts without it in mean and max tasks, especially for larger $N$. While $\tanh$ appears less beneficial in the sum task, real-world MARL scenarios are unlikely to involve purely additive interactions. Instead, joint effects often arise from a mixture of operations—such as averaging, maxima, and other nonlinear combinations—which $\tanh$ may help capture by introducing nonlinearity and bounded activation.

We believe mean and max are more meaningful when interpreting joint observations. For example, in many environments, position is a shared feature; its mean can represent the group's center of gravity, and the max indicates border positions. In contrast, the sum of positions is less informative.

Given its consistent second-best or best performance, the mean-pooling with $\tanh$ variant is a strong general-purpose candidate. To validate this further, we add a ``mix'' task:

\begin{equation*}
    y_i = x_i + \left((\sum_{j=1}^N x_j)/N + \sum_{j=1}^N x_j + \max_{j=1}^N x_j\right)/3
\end{equation*}

\begin{figure*}[ht]
    \centering
    \begin{subfigure}[b]{0.32\textwidth}
        \centering
        \includegraphics[width=\textwidth]{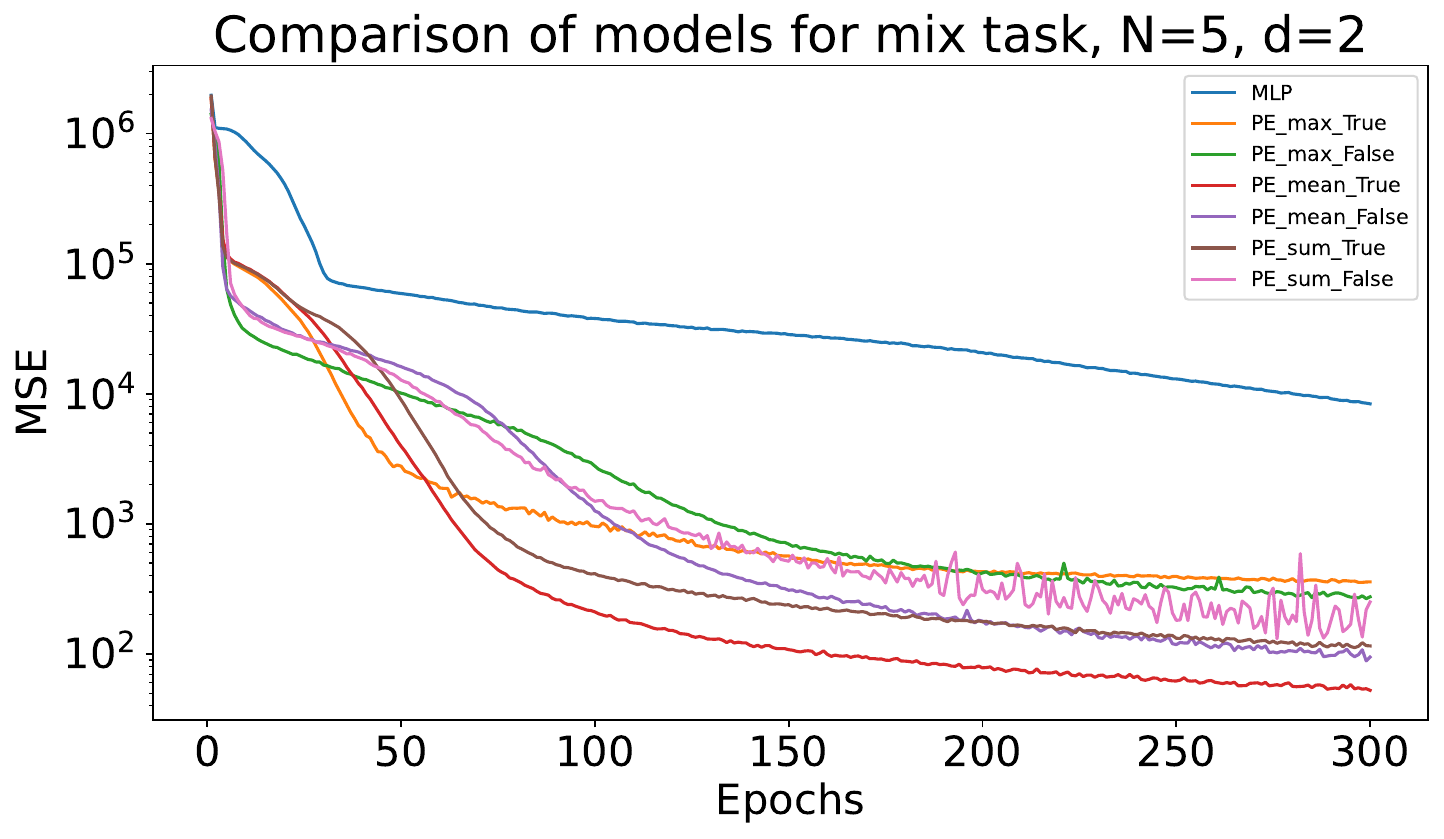}
        \caption{$N=5$}
        \label{fig:mix_2_5}
    \end{subfigure}
    \hfill
    \begin{subfigure}[b]{0.32\textwidth}
        \centering
        \includegraphics[width=\textwidth]{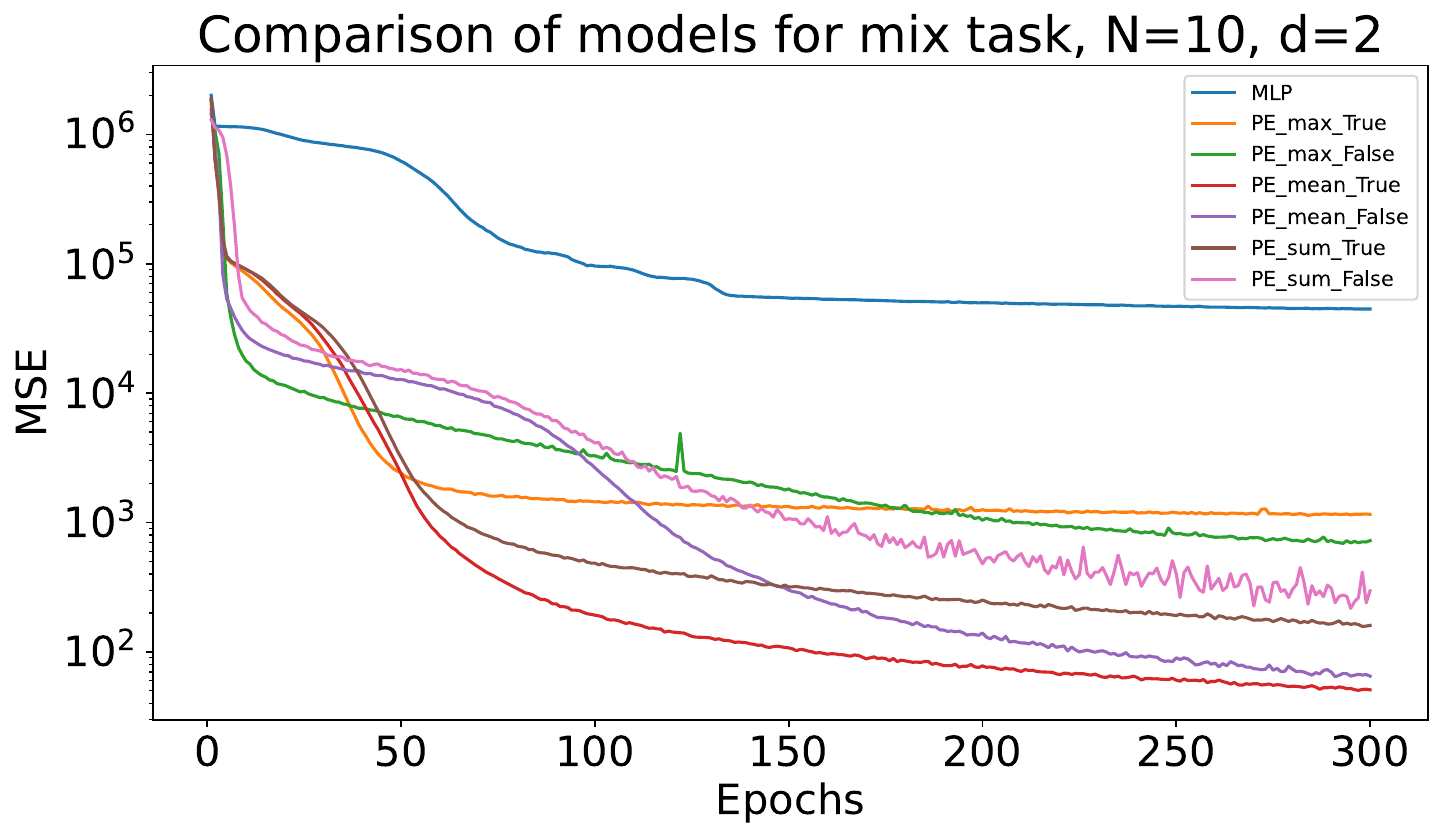}
        \caption{$N=10$}
        \label{fig:mix_2_10}
    \end{subfigure}
    \hfill
    \begin{subfigure}[b]{0.32\textwidth}
        \centering
        \includegraphics[width=\textwidth]{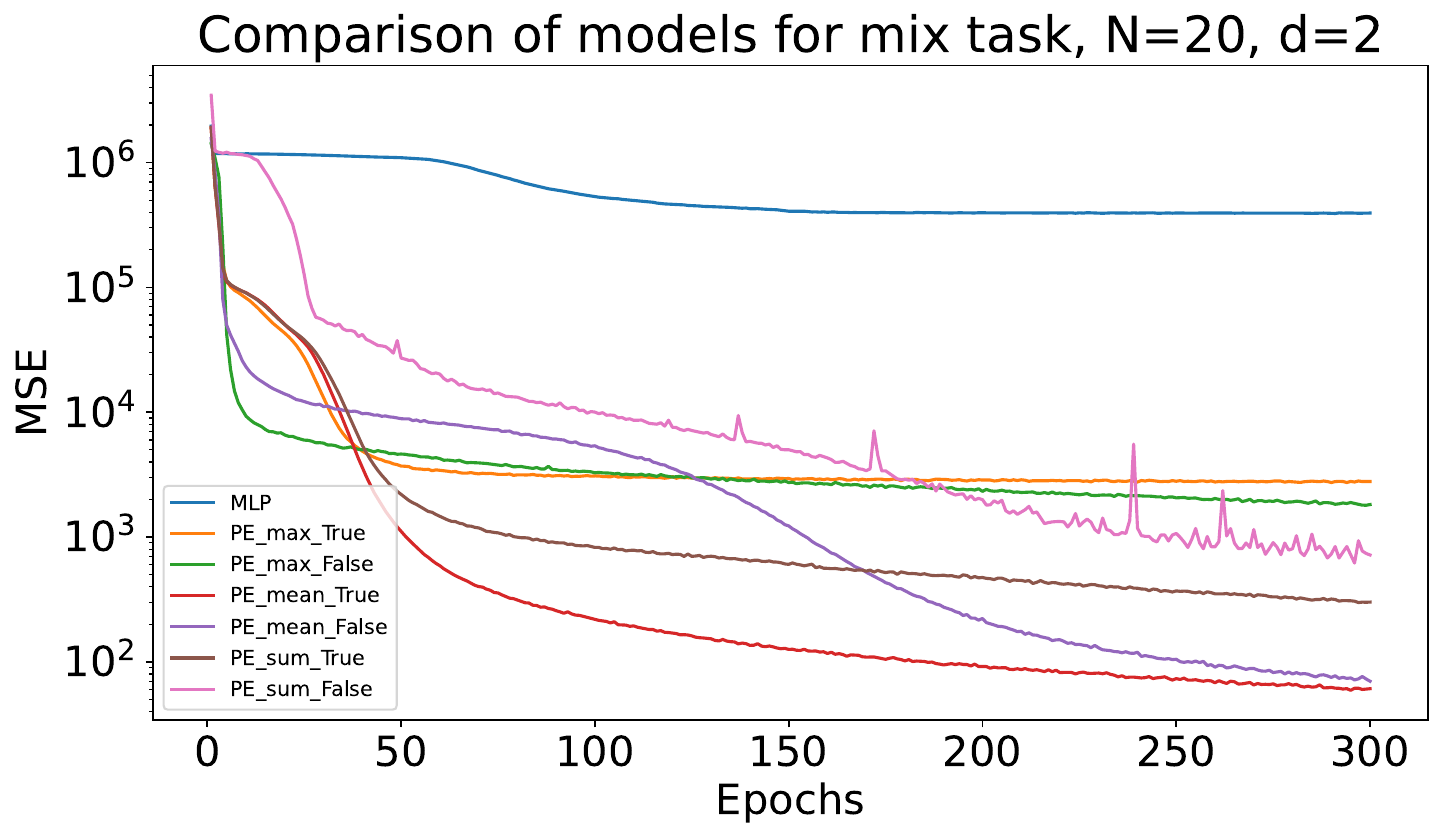}
        \caption{$N=20$}
        \label{fig:mix_2_20}
    \end{subfigure}
    \caption{Test on the mix task: $y_i = x_i + ((\sum_{j=1}^N x_j)/N + \sum_{j=1}^N x_j + \max_{j=1}^N x_j)/3$}
    \label{fig:toymix}
\end{figure*}

The results of this task (\Cref{fig:toymix}) confirm that mean pooling combined with $\tanh$ activation, precisely the configuration used in GLPE, achieves the best performance under hybrid joint influence patterns. This validates the scalability of our design and its ability to approximate complex joint effects in MARL, further justifying its use in the main experiments.

\section{Comparing the Size of \gls{glpe} and Centralized Policies}

\label{sec:glpe_param_count}
We analyze the parameter count of the \gls{glpe} policy introduced in~\Cref{sec:CPE}, comparing it to a standard distributed \gls{mlp} policy with the same architecture (i.e., identical layer count and hidden dimensions). Let $\left|\theta_{\text{MLP}}\right|$ denote the number of parameters in the \gls{mlp}. As stated in the main text, GLPE introduces only a modest increase in size. We formally show that the corresponding \gls{glpe} network contains at most twice the parameters:

\begin{equation}
\label{eq:nbparamglpe}
\left|\theta_{\text{GLPE}}\right| \leq 2 \cdot \left|\theta_{\text{MLP}}\right|.
\end{equation}

Consider a single \gls{mlp} layer $f_{\text{MLP}}: \mathbb{R}^d \to \mathbb{R}^{d'}$. The corresponding \gls{glpe} layer operates on the joint agent input $x \in \mathbb{R}^{N \times d}$ and maps it to $f_{\text{GLPE}}: \mathbb{R}^{N \times d} \to \mathbb{R}^{N \times d'}$. This layer consists of two components:
\begin{itemize}
    \item A local sub-layer $\sublayer_{\text{loc}}: \mathbb{R}^d \to \mathbb{R}^{d'}$ applied to each agent independently;
    \item A global sub-layer $\sublayer_{\text{glo}}: \mathbb{R}^{N \times d} \to \mathbb{R}^{d'}$ applied to the mean-pooled input across all agents.
\end{itemize}

The local sub-layer $\sublayer_{\text{loc}}$ has the same structure and parameter count as $f_{\text{MLP}}$. The global sub-layer is typically implemented as $\sublayer_{\text{glo}}(x) = \tanh\left(\sublayerfunc_{\text{pooling}}(\text{mean}(x))\right)$, where $\text{mean}(x) \in \mathbb{R}^d$ is the average input across agents, and $\sublayerfunc_{\text{pooling}}: \mathbb{R}^d \to \mathbb{R}^{d'}$ is a one-layer \gls{mlp}. Since $\tanh$ and $\text{mean}$ are parameter-free, the parameter count of $\sublayer_{\text{glo}}$ is entirely determined by $\sublayerfunc_{\text{pooling}}$, which again matches that of $f_{\text{MLP}}$. 

Thus, the total number of parameters in a \gls{glpe} layer is at most twice that of the corresponding \gls{mlp} layer, leading to~\eqref{eq:nbparamglpe}.

In practice, the actual parameter overhead can be smaller. For example, when using recurrent architectures such as GRUs, the global sub-layer is often omitted. Additionally, removing biases from the global sub-layer further reduces the parameter count, which explains the inequality in~\eqref{eq:nbparamglpe}.

This property guarantees that the parameter count of the \gls{glpe} policy scales linearly with that of the underlying \gls{mlp} policy and does not grow with the number of agents, making it suitable for large-scale multi-agent settings.

\section{Experiment Details and Additional Results}
\subsection{Environments}

We provide a detailed description of the environments used in our experiments, including their dynamics, individual observation $z_i$, and action space $A_i$. Note that the observation space depends on the configuration used during testing and may vary with different settings. For example, in SMAC, certain configuration options may lead to the inclusion or exclusion of specific observation features.

\textbf{MPE Spread.}  
\begin{figure}[H]
    \centering
    \includegraphics[width=0.4\linewidth]{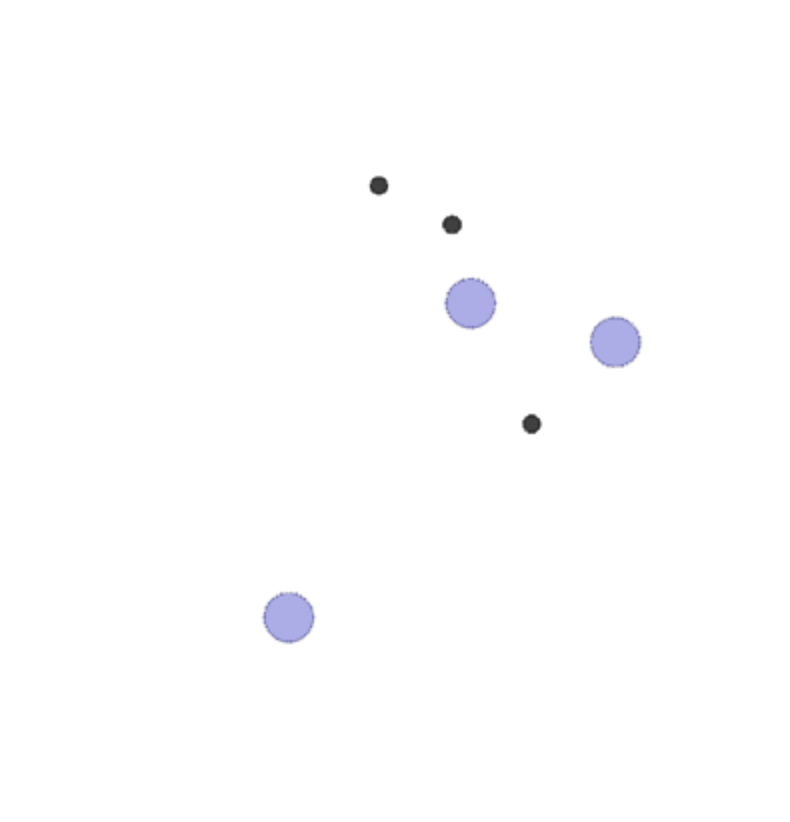}
    \caption{Spread task with $N=3$}
    \label{fig:spreadexample}
\end{figure}
The Multi-Agent Particle Environment (MPE) contains several cooperative tasks; among them, \textit{Spread}~(\Cref{fig:spreadexample}) requires $N$ agents to cover $N$ distinct landmarks while avoiding collisions. 
In the default setting, each agent observes the positions of all others. 
To increase the complexity and realism, specifically, to enforce partial observability and reduce communication, we disable other-agent observations. This version encourages implicit coordination without explicit awareness of teammates.

\textit{Individual observation:}
\begin{align*}
z_i = \big(&
\texttt{self\_velocity\_x},~ \texttt{self\_velocity\_y}, \\
& \texttt{self\_position\_x},~ \texttt{self\_position\_y}, \\
& (\texttt{relative\_position\_x\_landmark}_j,\\ 
& \texttt{relative\_position\_y\_landmark}_j)_{j=1}^N
\big)
\end{align*}

\textit{Individual action space:}
\begin{align*}
A_i = \{ \texttt{No move},~ \texttt{Left},~ \texttt{Right},~ \texttt{Down},~ \texttt{Up} \}
\end{align*}

\textit{Rewards:}  
All agents receive a shared global reward proportional to the negative sum of the minimum distances from each landmark to its nearest agent, encouraging full landmark coverage. Additionally, agents incur a local penalty of $-1$ for each collision with another agent, discouraging overlapping behaviors.

\begin{figure}
    \centering
    \includegraphics[width=0.6\linewidth]{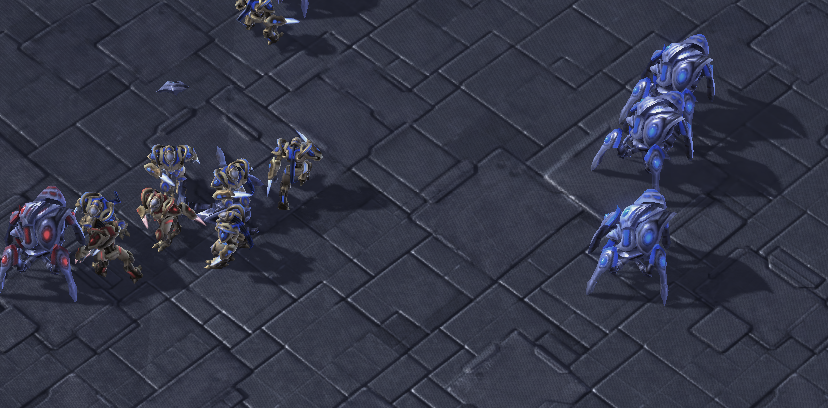}
    \caption{Screenshot of a SMAC scenario on the \texttt{3s5z\_vs\_3s6z} map.}
    \label{fig:smacexample}
\end{figure}

\textbf{SMAC.}  
The StarCraft Multi-Agent Challenge (SMAC)~(\Cref{fig:smacexample})~\cite{art:smac} is built on StarCraft II and features complex micromanagement scenarios in which each agent controls a single allied unit ($N$ agents in the allied team) against an enemy team ($M$ agents). It poses challenges such as heterogeneous unit types, long time horizons, and the need for strategic coordination. We evaluate performance across several standard maps that vary in difficulty and cooperation demands.

\textit{Individual observation:}
\begin{align}
z_i = \big(&
\texttt{avail\_move\_north},~ \texttt{avail\_move\_south}, \notag \\
& \texttt{avail\_move\_east},~ \texttt{avail\_move\_west}, \notag \\
& (\texttt{avail\_for\_attack}_j,~ \texttt{enemy\_distance}_j, \notag \\
& \quad \texttt{enemy\_relative\_x}_j,~ \texttt{enemy\_relative\_y}_j, \notag \\
& \quad \texttt{enemy\_health}_j,~ \texttt{enemy\_unit\_type}_j)_{j=1}^M, \notag \\
& (\texttt{visible}_j,~ \texttt{ally\_distance}_j, \notag \\
& \quad \texttt{ally\_relative\_x}_j,~ \texttt{ally\_relative\_y}_j, \notag \\
& \quad \texttt{ally\_health}_j,~ \texttt{ally\_unit\_type}_j, \notag \\
& \quad \texttt{ally\_last\_action}_j)_{j \in \{1,\ldots,N\} \setminus \{i\}}, \notag \\
& \texttt{self\_health},~ \texttt{self\_unit\_type}
\big) \notag
\end{align}

The features \texttt{self\_unit\_type}, \texttt{enemy\_unit\_type}$_j$, and \texttt{ally\_unit\_type}$_j$ are included only if multiple unit types are present on the map (e.g., not in \texttt{27m\_vs\_30m}). These fields are represented using one-hot encoding.

\textit{Individual action space:}
\begin{align*}
A_i =\; &\{ \texttt{No-Op},~ \texttt{Stop},~
\texttt{North},~ \texttt{South},~ \texttt{East},~ \texttt{West} \} \\
&\cup \{ \texttt{Attack enemy ID}_j \}_{j=1}^M
\end{align*}

\textit{Rewards:}  
A shaped reward function provides feedback based on the hit-point damage dealt to enemies, the number of enemies eliminated, and an additional bonus for winning the battle. The total shaped reward is normalized to a maximum of $20$.

\begin{figure}
    \centering
    \includegraphics[width=0.4\linewidth]{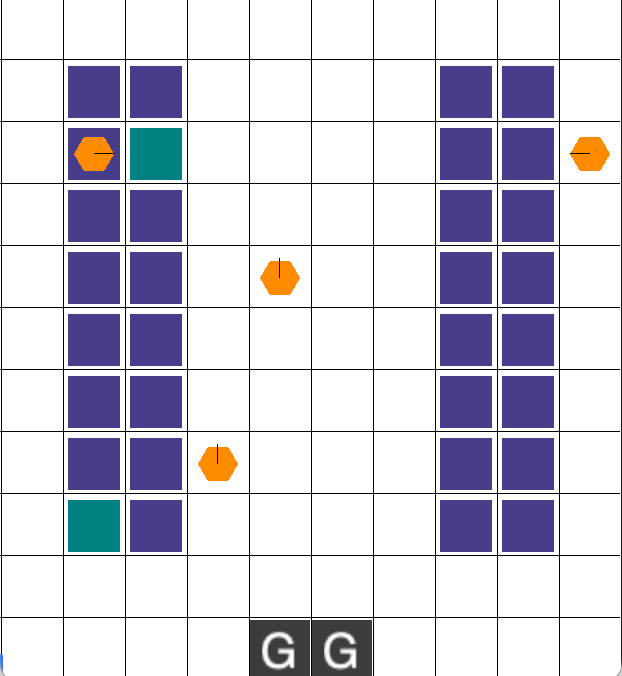}
    \caption{RWARE scenario on the tiny map with the hard difficulty setting.}
    \label{fig:rwareexample_appendix}
\end{figure}

\textbf{RWARE.}  
The Multi-Robot Warehouse Environment (RWARE)~(\Cref{fig:rwareexample_appendix}) simulates a warehouse where agents must retrieve and deliver requested shelves. Agents operate in a grid world with partial observability, sparse rewards, and limited communication. We adopt the ``hard'' configuration, which requests less shelves simultaneously, emphasizing coordinated planning and collision avoidance.

\textit{Individual observation:}
\begin{align*}
z_i = \big(&
\texttt{self\_position\_x},~ \texttt{self\_position\_y}, \\
& \texttt{carrying\_shelf},~ \texttt{direction\_one\_hot}, \\
& \texttt{path\_restricted}, \\
& (\texttt{is\_agent}_{(j,k)}, \\
& \quad \texttt{agent\_direction}_{(j,k)},~ \texttt{is\_shelf}_{(j,k)}, \\
& \quad \texttt{is\_shelf\_required}_{(j,k)})_{(j, k) \in \{1,2,3\} \times \{1,2,3\}}
\big)
\end{align*}

Here, the features \texttt{direction\_one\_hot} and \texttt{agent\_direction}$_{(j,k)}$ are one-hot encodings over four directions: up, down, left, and right. The boolean \texttt{path\_restricted} indicates whether the agent is in a zone where it cannot carry a shelf. The flattened $3 \times 3$ grid ($(\dots)_{(j,k)}$) represents the agent's local surroundings.

\textit{Individual action space:}
\begin{align*}
A_i = \{&
\texttt{No move},~ \texttt{Move forward}, \\
& \texttt{Turn left},~ \texttt{Turn right}, \\
& \texttt{Load or unload shelf}
\}
\end{align*}

\textit{Rewards:}  
An agent receives a reward of $1$ when it successfully delivers a requested shelf. Since delivery requires a sequence of coordinated actions and rewards are only issued upon successful completion, the reward signal is inherently sparse.

\subsection{Best Policy Results}
\begin{figure*}
    \centering
    \includegraphics[width=0.9\linewidth]{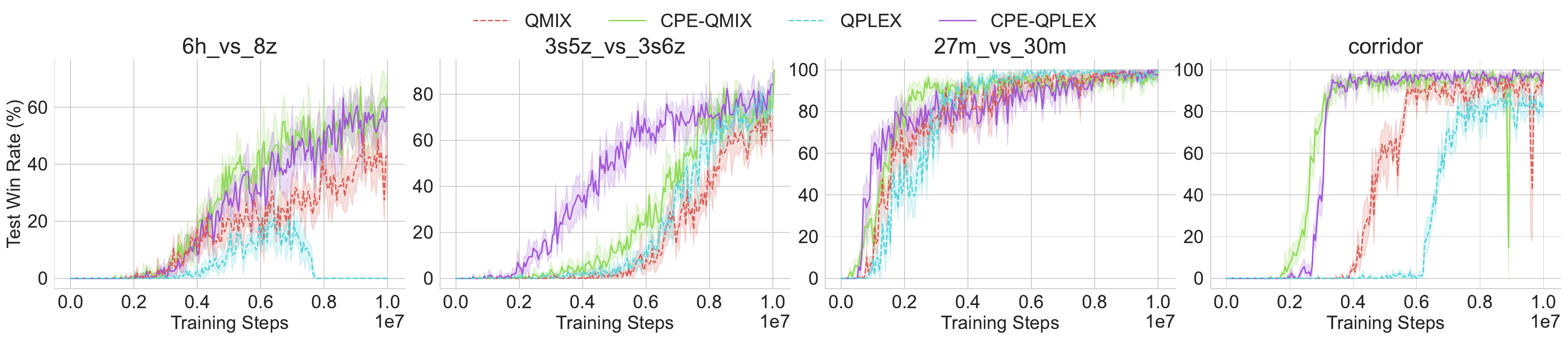}
    \caption{Training curve of the best policy (highest final test performance) on \gls{smac} maps.}
    \label{fig:smactop}
\end{figure*}

\begin{figure*}
    \centering
    \includegraphics[width=0.9\linewidth]{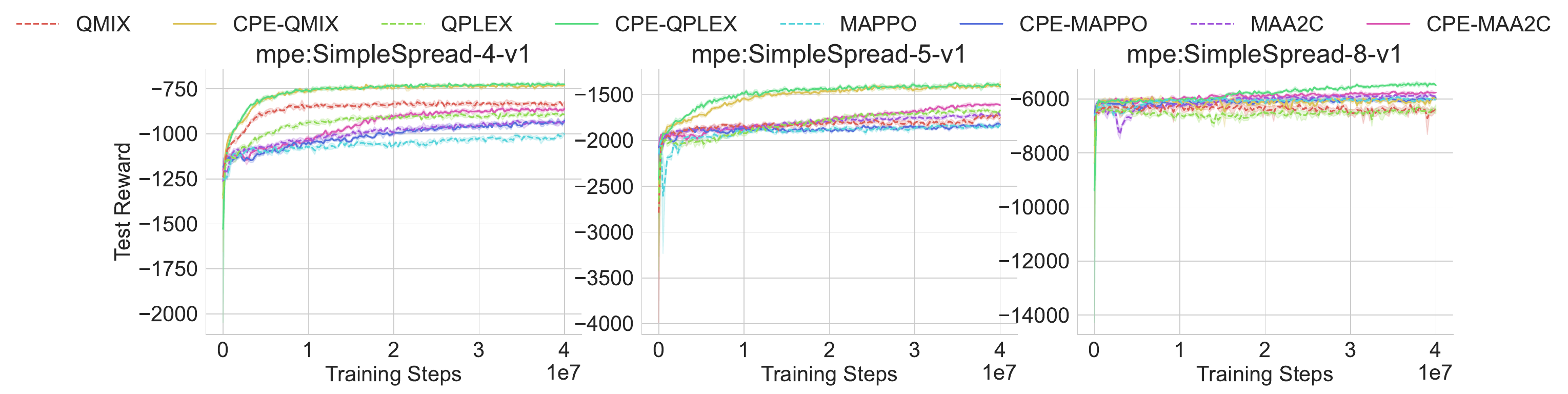}
    \caption{Training curves of the best policies on \gls{mpe} benchmarks.}
    \label{fig:mpetop}
\end{figure*}

\begin{figure*}
    \centering
    \includegraphics[width=0.9\linewidth]{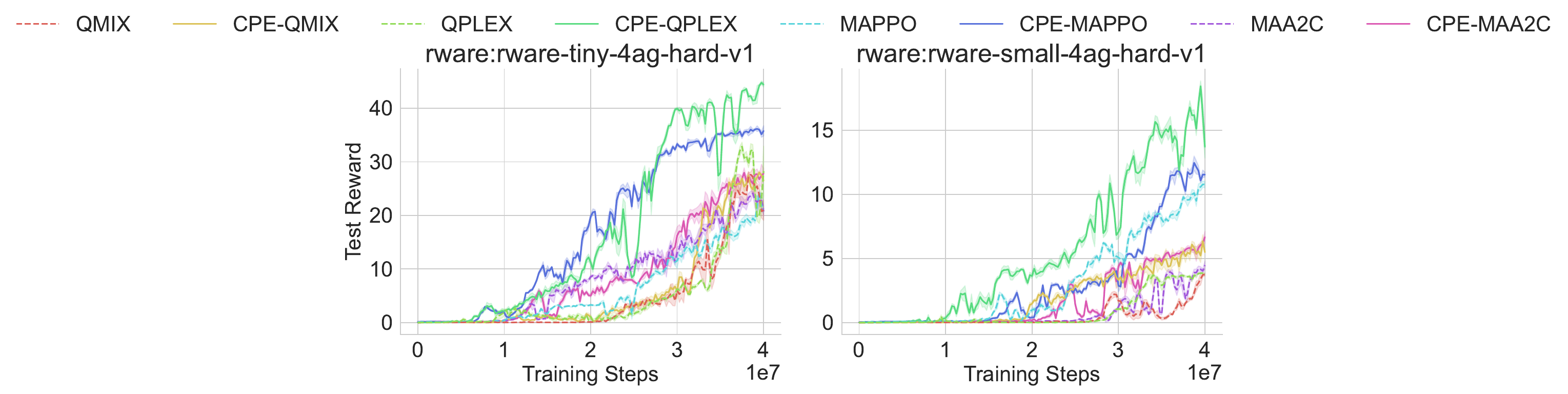}
    \caption{Training curves of the best policies on \gls{rware}.}
    \label{fig:rwaretop}
\end{figure*}

We report the performance of the best policy, measured by the average win rate (for \gls{smac}) or test episodic reward (for \gls{mpe} and \gls{rware}) over the final 10 evaluation rounds across five random seeds. The results are shown in~\Cref{fig:smactop,fig:mpetop,fig:rwaretop} and~\Cref{tab:smactop,tab:topmpe,tab:toprware}.

Overall, the findings are consistent with trends observed in mean performance, indicating that the benefits of CPE extend beyond average outcomes to best-case scenarios. This reinforces the robustness of our centralized approach. Additionally, the results suggest that CPE’s advantage stems not only from access to global information, but also from the effective utilization of that information via the \gls{glpe} architecture, even in settings like \gls{smac}, where observations are already relatively rich. Notably, for algorithms with higher variance, such as QPLEX and CPE-QPLEX, best-case performance often surpasses the mean, further highlighting the potential of our approach. 

\begin{table*}[hbt]
    \centering
    \begin{tabular}{lcccc}
        \toprule
        \textbf{Map} & \textbf{CPE-QMIX} & \textbf{CPE-QPLEX} & \textbf{QMIX} & \textbf{QPLEX} \\
        \midrule
        6h\_vs\_8z & 61.25 & 57.50 & 42.81 & 0.00 \\
        3s5z\_vs\_3s6z & 75.94 & 84.38 & 65.94 & 77.50 \\
        27m\_vs\_30m & 97.81 & 97.50 & 99.38 & 99.38 \\
        corridor & 98.12 & 96.25 & 93.12 & 82.50 \\
        \bottomrule
    \end{tabular}
    \caption{Final test performance of the best policies (highest average over last 10 test episodes) on \gls{smac}.}
    \label{tab:smactop}
\end{table*}

\begin{table*}[hbt]
    \centering
    \begin{tabular}{lcccc}
        \toprule
        \textbf{Map} & \textbf{QMIX} & \textbf{CPE-QMIX} & \textbf{QPLEX} & \textbf{CPE-QPLEX} \\
        \midrule
        SimpleSpread-4 & $-833.18 \pm 117.7$ & $-732.77 \pm 88.88$ & $-893.30 \pm 140.11$ & $-724.54 \pm 103.48$ \\
        SimpleSpread-5 & $-1731.78 \pm 237.57$ & $-1404.44 \pm 226.95$ & $-1694.48 \pm 235.81$ & $-1392.41 \pm 237.41$ \\
        SimpleSpread-8 & $-6424.75 \pm 971.32$ & $-5991.82 \pm 780.97$ & $-6423.85 \pm 901.11$ & $-5465.61 \pm 666.03$ \\
        \bottomrule
    \end{tabular}
    
    \vspace{0.5em}

    \begin{tabular}{lcccc}
        \toprule
        \textbf{Map} & \textbf{MAPPO} & \textbf{CPE-MAPPO} & \textbf{MAA2C} & \textbf{CPE-MAA2C} \\
        \midrule
        SimpleSpread-4 & $-1005.38 \pm 133.34$ & $-939.60 \pm 131.11$ & $-935.86 \pm 127.33$ & $-866.68 \pm 120.84$ \\
        SimpleSpread-5 & $-1844.42 \pm 212.36$ & $-1816.13 \pm 205.44$ & $-1709.71 \pm 207.84$ & $-1605.05 \pm 186.66$ \\
        SimpleSpread-8 & $-6026.01 \pm 590.21$ & $-5958.40 \pm 826.70$ & $-5894.33 \pm 540.97$ & $-5750.41 \pm 681.28$ \\
        \bottomrule
    \end{tabular}
    
    \caption{Final test episodic rewards of the best policies on \gls{mpe}.}
    \label{tab:topmpe}
\end{table*}

\begin{table*}[hbt]
    \centering
    \begin{tabular}{llcccc}
        \toprule
        \textbf{Map} & \textbf{QMIX} & \textbf{CPE-QMIX} & \textbf{QPLEX} & \textbf{CPE-QPLEX} \\
        \midrule
        tiny-4ag-hard & $21.52 \pm 6.18$ & $27.41 \pm 4.05$ & $23.20 \pm 7.20$ & $44.57 \pm 2.90$ \\
        small-4ag-hard & $3.64 \pm 0.76$ & $5.92 \pm 2.25$ & $3.84 \pm 0.63$ & $16.45 \pm 2.96$ \\
        \bottomrule
    \end{tabular}
    
    \vspace{0.5em}

    \begin{tabular}{llcccc}
        \toprule
        \textbf{Map} & \textbf{MAPPO} & \textbf{CPE-MAPPO} & \textbf{MAA2C} & \textbf{CPE-MAA2C} \\
        \midrule
        tiny-4ag-hard & $22.43 \pm 5.52$ & $35.49 \pm 4.73$ & $22.41 \pm 5.14$ & $27.27 \pm 3.73$ \\
        small-4ag-hard & $10.70 \pm 2.54$ & $11.36 \pm 2.86$ & $4.11 \pm 1.21$ & $6.25 \pm 1.51$ \\
        \bottomrule
    \end{tabular}

    \caption{Final test episodic rewards of the best policies on \gls{rware}.}
    \label{tab:toprware}
\end{table*}

\subsection{Training Time}
\label{sec:traintime}
A potential concern with centralized policies is computational efficiency. Unlike distributed networks, which operate on individual observations and typically process 1D inputs of shape $[|z_i|]$, centralized policies must handle joint observations. While the \gls{glpe} network is both lightweight and agent-number agnostic, it requires input tensors in 2D form ($[N, |z_i|]$). For a training batch of size $bs$, distributed policies can flatten the input into shape $[bs \cdot N, |z_i|]$, allowing efficient parallelization on GPUs. In contrast, \gls{glpe} inputs must retain the shape $[bs, N, |z_i|]$, which limits the effectiveness of batch processing. Additionally, the global sub-layer in \gls{glpe} introduces extra computation compared to purely local distributed networks.

To quantify the impact, we measured the training time of CPE-enhanced algorithms versus their CTDE counterparts. \Cref{tab:smac_traintime} reports the average wall-clock training time across five random seeds for four selected \gls{smac} maps.

All experiments were conducted on a single compute node equipped with an NVIDIA A100 GPU and 16 physical CPU cores (with hyperthreading disabled). Each run was allocated one GPU and had exclusive access to the node. All algorithms were trained using the parallel runner provided in the PyMARL framework.

\begin{table*}[t]
    \centering
    \begin{tabular}{lcccccc}
        \toprule
        \textbf{Map} & \textbf{QMIX} & \textbf{CPE-QMIX} & \textbf{Increase} & \textbf{QPLEX} & \textbf{CPE-QPLEX} & \textbf{Increase} \\
        \midrule
        6h\_vs\_8z & 4h13min & 4h39min & 10.28\% & 4h00min & 4h36min & 15.00\% \\
        3s5z\_vs\_3s6z & 4h51min & 5h15min & 8.25\% & 4h54min & 5h23min & 9.86\% \\
        27m\_vs\_30m & 15h14min & 16h18min & 7.00\% & 16h24min & 19h56min & 21.54\% \\
        corridor & 8h08min & 8h40min & 6.56\% & 8h08min & 9h15min & 13.73\% \\
        \bottomrule
    \end{tabular}
    \caption{Training time for distributed and \gls{glpe}-based policies across selected \gls{smac} maps.}
    \label{tab:smac_traintime}
\end{table*}

Overall, CPE introduces only a moderate training time overhead, ranging from 6.56\% to 21.54\% across different maps and algorithms. QPLEX generally experiences a higher increase than QMIX, while the effect of the map or number of units appears less consistent. Notably, in some cases (e.g., 27m\_vs\_30m), the difference between CPE and its base method is smaller than the difference between two CTDE baselines (QMIX vs. QPLEX). These findings show that CPE’s added training cost is acceptable given its performance gains, easing concerns about scalability in practice.

\subsection{Hyperparameter Settings}
We report the hyperparameters used for each algorithm across different environments. Most configurations follow the settings from~\cite{art:hpn} for \gls{smac} and~\cite{art:pymarlzooplus} for \gls{mpe} and \gls{rware}. All CPE-augmented algorithms share the same hyperparameters as their baseline counterparts to ensure fair comparison, except that vanilla \gls{ctde} methods use GRU-based networks, while their CPE variants employ GLPE-GRU architectures.

\begin{table}[h]
\centering
\begin{tabular}{ll}
\toprule
\textbf{Name} & \textbf{Value} \\
\midrule
agent runner & parallel(8) \\
optimizer & Adam \\
batch size & 128 \\
hidden dimension & 64 \\
learning rate & 0.001 \\
network type & GRU / GLPE-GRU \\
epsilon anneal & 100000 \\
epsilon start & 1.0 \\
epsilon finish & 0.05 \\
target update & 200 \\
buffer size & 5000 \\
$\gamma$ (discount factor) & 0.99 \\
observation agent id & True \\
observation last action & False \\
mixing network hidden dimension & 32 \\
hypernetwork dimension & 64 \\
\bottomrule
\end{tabular}
\caption{Hyperparameters for QMIX and CPE-QMIX for SMAC}
\label{tab:hpqmix_smac}
\end{table}

\begin{table}[h]
\centering
\begin{tabular}{ll}
\toprule
\textbf{Name} & \textbf{Value} \\
\midrule
agent runner & parallel(8) \\
optimizer & Adam \\
batch size & 128 \\
hidden dimension & 64 \\
learning rate & 0.001 \\
network type & GRU / GLPE-GRU \\
epsilon anneal & 100000 \\
epsilon start & 1.0 \\
epsilon finish & 0.05 \\
target update & 200 \\
buffer size & 5000 \\
$\gamma$ (discount factor) & 0.99 \\
observation agent id & True \\
observation last action & False \\
mixing network hidden dimension & 32 \\
hypernetwork dimension & 64 \\
\bottomrule
\end{tabular}
\caption{Hyperparameters for QPLEX and CPE-QPLEX for SMAC}
\label{tab:hpqplex_smac}
\end{table}

\begin{table}[h]
\centering
\begin{tabular}{ll}
\toprule
\textbf{Name} & \textbf{Value} \\
\midrule
agent runner & episode \\
optimizer & Adam \\
batch size & 32 \\
hidden dimension & 64 (MPE)/128 (RWARE) \\
learning rate & 0.0005 \\
reward standardisation & True \\
network type & GRU / GLPE-GRU \\
evaluation epsilon & 0.0 \\
epsilon anneal & 50000 \\
epsilon start & 1.0 \\
epsilon finish & 0.05 \\
target update & 200 \\
buffer size & 5000 \\
$\gamma$ (discount factor) & 0.99 \\
observation agent id & True \\
observation last action & True \\
mixing network hidden dimension & 32 \\
hypernetwork dimension & 64 \\
hypernetwork number of layers & 2 \\
\bottomrule
\end{tabular}
\caption{Hyperparameters for QMIX and CPE-QMIX for MPE and RWARE}
\label{tab:hpqmix}
\end{table}

\begin{table}[h]
\centering
\begin{tabular}{ll}
\toprule
\textbf{Name} & \textbf{Value} \\
\midrule
agent runner & episode \\
optimizer & RMSProp \\
batch size & 32 \\
hidden dimension & 64 (MPE)/128 (RWARE) \\
learning rate & 0.0005 \\
reward standardisation & True \\
network type & GRU / GLPE-GRU \\
evaluation epsilon & 0.0 \\
epsilon anneal & 200000 \\
epsilon start & 1.0 \\
epsilon finish & 0.05 \\
target update & 200 \\
buffer size & 5000 \\
$\gamma$ (discount factor) & 0.99 \\
observation agent id & True \\
observation last action & True \\
mixing network hidden dimension & 32 \\
hypernetwork dimension & 64 \\
hypernetwork number of layers & 2 \\
\bottomrule
\end{tabular}
\caption{Hyperparameters for QPLEX and CPE-QPLEX for MPE and RWARE}
\label{tab:hpqplex}
\end{table}

\begin{table}[h]
\centering
\begin{tabular}{ll}
\toprule
\textbf{Name} & \textbf{Value} \\
\midrule
agent runner & parallel(10) \\
optimizer & Adam \\
batch size & 10 \\
hidden dimension & 64 (MPE)/128 (RWARE) \\
learning rate & 0.0005 \\
reward standardisation & True \\
network type & GRU / GLPE-GRU \\
entropy coefficient & 0.01 \\
target update & 200 \\
buffer size & 10 \\
$\gamma$ (discount factor) & 0.99 \\
observation agent id & True \\
observation last action & True \\
n-step & 5 \\
epochs & 4 \\
clip & 0.2 \\
\bottomrule
\end{tabular}
\caption{Hyperparameters for MAPPO and CPE-MAPPO for MPE and RWARE}

\label{tab:hpmappo}
\end{table}

\begin{table}[h]
\centering

\begin{tabular}{ll}
\toprule
\textbf{Name} & \textbf{Value} \\
\midrule
agent runner & parallel(10) \\
optimizer & Adam \\
batch size & 10 \\
hidden dimension & 64 (MPE)/128 (RWARE) \\
learning rate & 0.0005 \\
reward standardisation & True \\
network type & GRU / GLPE-GRU \\
entropy coefficient & 0.01 \\
target update & 200 \\
buffer size & 10 \\
$\gamma$ (discount factor) & 0.99 \\
observation agent id & True \\
observation last action & True \\
n-step & 5 \\
\bottomrule
\end{tabular}
\caption{Hyperparameters for MAA2C and CPE-MAA2C for MPE and RWARE}
\label{tab:hpmaa2c}
\end{table}

\subsection{Code}
The code for our experiments is included in the supplementary material and will be made publicly available upon acceptance of this work.

\section{Discussion on the Limitations of Centralized Policies}

While centralized policies offer clear advantages in coordination, they also face practical limitations in real-world settings—particularly with respect to communication and synchronization, which are often underrepresented in mainstream MARL benchmarks. 
However, as shown in our experiments on \gls{mpe}, some environments implicitly rely on information that would, in realistic settings, require communication to access.
This raises questions about the robustness of distributed policies under real-world constraints such as partial observability and limited inter-agent communication.

In our experiments, we did not observe a substantial increase in training time due to CPE, as detailed in~\Cref{sec:traintime}. 
While centralized policies require more communication bandwidth, they benefit from centralized computation, avoiding the need to deploy networks on individual agents.

Ultimately, our work challenges the common assumption that distributed policies always lead to better policy quality. We propose a centralized alternative that excels in scenarios where coordination is essential and centralized computation is feasible. We also advocate for the development of benchmarks that more accurately capture real-world constraints, such as limited communication or restricted local computation. While centralized execution may encounter practical limitations beyond current benchmarks, our findings demonstrate that, when these constraints are managed, centralized policies can offer substantial coordination benefits—an aspect we believe the community should not overlook.

\end{document}